\documentclass{emulateapj}
\usepackage{lscape}

\shorttitle{Radial Velocity TATOOINE Search for Circumbinary Planets}

\shortauthors{M. Konacki, M. Muterspaugh, S. Kulkarni and K. Helminiak}

\begin{document}

\title{The Radial Velocity TATOOINE Search for Circumbinary Planets: Planet 
Detection Limits for a Sample of Double-lined Binary Stars --- Initial Results 
from Keck I/Hires, Shane/CAT/Hamspec and TNG/Sarg Observations}

\author{Maciej Konacki\altaffilmark{1,2,3}, Matthew W.
Muterspaugh\altaffilmark{4,5}, Shrinivas R. Kulkarni\altaffilmark{6}
and Krzysztof G. He{\l}miniak\altaffilmark{1}}

\altaffiltext{1}{Nicolaus Copernicus Astronomical Center, Polish Academy of
Sciences, Rabianska 8, 87-100 Torun, Poland}
\altaffiltext{2}{Astronomical Observatory, A. Mickiewicz University,
Sloneczna 36, 60-286 Poznan, Poland}
\altaffiltext{3}{e-mail: maciej@ncac.torun.pl}
\altaffiltext{4}{Department of Mathematics and Physics, College of Arts
and Sciences, Tennessee State University, Boswell Science Hall,
Nashville, TN 37209, USA}
\altaffiltext{5}{Tennessee State University, Center of Excellence in Information
Systems, 3500 John A. Merritt Blvd., Box No. 9501,
Nashville, TN 37203-3401, USA}
\altaffiltext{6}{Division of Physics, Mathematics and Astronomy, California Institute
of Technology, Pasadena, CA 91125, USA}

\begin{abstract}
We present preliminary results of the first and on-going radial velocity survey for 
circumbinary planets. With a novel radial velocity technique employing an iodine 
absorption cell we achieve an unprecedented RV precision of up to 2 m$\,$s$^{-1}$
for double-lined binary stars. The high resolution spectra collected with the Keck 
I/Hires, TNG/Sarg and Shane/CAT/Hamspec telescopes/spectrographs over the years 
2003-2008 allow us to derive RVs and compute planet detection limits for ten double-lined
binary stars. For this initial sample of targets, we can rule out planets on dynamically 
stable orbits with masses as small as $\sim$0.3 to 3 M$_{Jup}$ for the orbital periods 
of up to $\sim$5.3 years. Even though the presented sample of stars is too small to make 
any strong conclusions, it is clear that the search for circumbinary planets is now 
technique-wise possible and eventually will provide new constraints for the planet 
formation theories.
\end{abstract}


\section{Introduction}

Searches for planets in close binary systems explore the degree to which
stellar multiplicity inhibits or promotes planet formation
\citep{Mut:05::,Mut:06::,Mut:07::}. 
Detection of giant planets orbiting both components of
short period ($P < 60$ days) binaries (``circumbinary planets'') will have
significant consequences for theoretical understandings of how giant 
planets are formed.  The binarity of the central body creates an environment
in which the evolution of a protoplanetary disk is substantially different 
than around single stars \citep{Artymowicz:94::}. 
This must have an effect on the migration of giant planets in a disk   
as well as on the ``parking'' mechanism and their final orbit. Likely, also    
the dynamical interaction between protoplanets and then planets in
a multi-planet system should be affected by the central body binarity
and presumably result in a different distribution of the orbital elements
of planets. Finally, if one assumes that planetary orbits are coplanar with
the orbit of an eclipsing binary, then there is an enhanced probability of 
detecting a circumbinary transiting planet
\citep{Konacki:09::,Ofir:09::,Ofir:08::,Schn:94::,Schn:90::}. 

The recent work by Konacki (2005,2009) demonstrates a method for obtaining 
radial velocity (RV) precisions of up to $5\,{\rm m\,s^{-1}}$ for double-lined 
spectroscopic binaries (SB2s) (now improved to 2 m$\,$s$^{-1}$) and for the first 
time opens  the opportunity to search for circumbinary planets via RVs. This method 
has been applied to a new search for circumbinary planets: The Attempt To Observe 
Outer-planets In Non-single-stellar Environments (TATOOINE).

Planets in binary and multiple stars have been a subject of theoretical
works long before any extrasolar planets were detected. \cite{Dvorak:82::}
investigated dynamical stability of a wide range of planetary
configurations in the framework of the elliptic restricted three-body  
problem. These included the so-called P-type (Planet-type, 
circumbinary orbits), S-type (Satellite-type, circumprimary or 
circumsecondary orbits) and L-type orbits (Librator-type, orbits 
around stable Lagrangian points L4 or L5 for the mass ratios $\mu<0.04$).
This subject was continued by e.g. \cite{Holman:99::} who 
based on extensive numerical simulations provided useful formulas  
allowing one to compute stable regions for among others circumbinary planets.

Circumbinary disks and planet formation in such disks have been of some
interest to theorists as well. The evolution of a circumbinary disk is 
studied by \cite{Artymowicz:94::} who demonstrates that such a disk will 
be truncated at its inner edge by tidal torques to within 1.8-2.6 times the semimajor 
axis of the binary. \cite{Moriwaki:04::} and \cite{Scholl:07::} show that 
planetessimal accretion should be possible in circumbinary disks and \cite{Quintana:06::} 
provide numerical proofs that planetary systems similar to those around single 
stars may be formed around binary stars with the apastron distance $\le$ 0.2 AU. 
Most recently, in a series of papers \cite{Pierens:07::,Pierens:08a::,Pierens:08b::} 
investigate formation, migration and evolution of planets in circumbinary
disks. In particular, they suggest that circumbinary planets may be
more common in the Saturn-mass regime as higher mass planets are more likely
to undergo close encounters with the secondary star \citep{Pierens:08b::}.

There are known a few cases of young spectroscopic binaries with
circumbinary disks. These include AK Sco \citep{Andersen:89::}, GW Ori
\citep{Mathieu:91::}, DQ Tau \citep{Mathieu:94::}, GG Tau \citep{Dutrey:94::}. 
The observations confirm the truncation of the disks at their inner edge
as predicted by theory. Recently, \cite{Kastner:08::} have detected CO, HCN,
CN and HCO$^{+}$ around a $\sim 12$ Myr-old close binary V4046 Sgr
demonstrating that it is surrounded by a rich molecular disk and showing 
a great similarity to the $\sim 8$ Myr-old star TW Hya; a star/disk system 
regarded a representative of the early solar nebula. Finally, \cite{Z:08::}
provide evidence that an SB2 BD+20 307 is an old binary star surrounded by 
a debris disk likely formed in a collision involving a terrestial planet.

Clearly, there is enough evidence that circumbinary planets should form,
evolve and survive on stable orbits around close binary stars. Yet no
radial velocity survey has been carried out to detect such planets despite
the fact that the RV technique for single stars has allowed for a thriving
scientific endeavor over the last 15 years. The fundamental problem with 
double-lined spectroscopic binary stars is that their spectra are highly 
variable due to the orbital motion of their components resulting in 
Doppler shifts typically up to $\sim$100 km$\,$s$^{-1}$ for each component spectrum.
Hence, the approach used for single stars where a Doppler shift of an
otherwise constant shape (spectrum) is found cannot be applied.

In this paper we present the preliminary results of the first radial velocity
survey for circumbinary planets. In \S{2} we discuss the novel
iodine cell based approach that allows us to measure precisely RVs
of SB2s. In \S{3} we describe in more detail our on-going RV effort
to detect circumbinary planets. In \S{4} we show the planet detection
limits for 10 SB2s from our survey and conclude in \S{5}.

\section{Precision RVs of double-lined spectroscopic binaries}

In the iodine cell (I$_{2}$) technique, the Doppler shift of a star
spectrum $\Delta\lambda$ is determined by solving the following
equation \citep{Mar:92::}.
\begin{equation}
\label{i2::}
I_{obs}(\lambda) =
[I_{s}(\lambda+\Delta\lambda_{s})\,T_{I_{2}}(\lambda+\Delta\lambda_{I_{2}})]
\,\otimes\,PSF
\end{equation}
where $\Delta\lambda_{s}$ is the shift of the star spectrum,
$\Delta\lambda_{I_{2}}$ is the shift of the iodine transmission
function $T_{I_{2}}$, $\otimes$ represents a convolution and   
$PSF$ a spectrograph's point spread function. The parameters 
$\Delta\lambda_{s}, \Delta\lambda_{I_{2}}$ as well as parameters 
describing the PSF are determined by performing a least-squares 
fit to the observed (through the iodine cell) spectrum $I_{obs}$. 
For this purpose, one also needs (1) a high signal to noise ratio 
(SNR) star spectrum taken  without the cell $I_{s}$ which serves 
as a template for all the spectra observed through the cell and 
(2) the I$_2$ transmission function $T_{I_{2}}$ obtained, for example, 
with the Fourier Transform Spectrometer at the Kitt Peak National Observatory.
The Doppler shift of a star spectrum is then given by
$\Delta\lambda = \Delta\lambda_{s} - \Delta\lambda_{I_{2}}$.
Such an iodine technique can only be applied to single stars. 
This is dictated by the need to supply an observed template spectrum 
of each component of a target binary star in Eq. \ref{i2::}. 

In the case when a composite spectrum of a binary star is observed, the
classical approach with the iodine cell cannot be used since it is
not possible to observationally obtain two separate template spectra
of the binary components.  This problem can be resolved as follows.
First, two sequential exposures of each (binary) target are always taken---one
with and the other without the cell.  This way one obtains an instantaneous
template that is used to model only the adjacent exposure taken with the
cell. The derived Doppler shift, $\Delta\lambda_i$ (where $i$ denotes the epoch of 
the observation), carries no meaning since each time a different template is 
used.  Moreover, it describes a Doppler ``shift'' of a composed spectrum that 
is typically different at each epoch. However, the parameters---in particular 
the wavelength solution and the parameters describing PSF---are accurately 
determined and can be used to extract the star spectrum, 
$I^{\star,i}_{obs}(\lambda)$, for each epoch $i$: 
\begin{equation}
\label{met::}
I^{\star,i}_{obs}(\lambda) = [I^{i}_{obs}(\lambda)\,\otimes^{-1}\,PSF^{i}]
/T_{I_{2}}(\lambda),
\end{equation}
where $\otimes^{-1}$ denotes deconvolution, and $PSF^{i}$ represents the
set of parameters describing PSF at the epoch $i$.  Such a star spectrum has
an accurate wavelength solution, is free of the I$_2$ lines and the 
influence of a varying PSF. 

In the next step, the velocities of both components of a binary target are
measured with the well know two-dimensional cross-correlation technique
TODCOR \citep{Zuc:94::} using as templates the synthetic spectra derived
with the ATLAS~9 and ATLAS~12 programs \citep{Kurucz:95::} and matched to 
the observed spectrum, $I_{s}(\lambda)$. This approach results in an RV
precision of 20-30 m$\,$s$^{-1}$ \citep{Konacki:05a::}. Now, having the RVs obtained 
with TODCOR, one can carry out a tomographic disentangling of the composite 
spectra of a binary to produce the real (observed) component spectra and
use them in the original equation \ref{i2::}. This finally allows us to
directly measure the RVs without relying on synthetic templates. Such a procedure 
results in RVs having a precision of up to several m$\,$s$^{-1}$ \citep{Konacki:09::}. 
The formal errors of the velocities are derived from the scatter between the 
velocities from different echelle orders.

\section{The TATOOINE survey}

The TATOOINE survey was initiated in mid 2003 with the 10-m Keck I and its Hires 
spectrograph \citep{Vogt:94::} as an addition to an RV survey for planets orbiting components
of speckle binary and multiple stars \citep{Konacki:05b::}. The original sample of 12 
SB2s was selected primarily to provide precision RVs for the astrometric data
collected at the Palomar Testbed Interferometer \citep{Colavita:99::,Konacki:04::}
and to test the new RV technique on a sample of close binaries. The Keck
I/Hires survey was continued until mid 2007. The sample was also monitored 
from mid 2006 until mid 2007 with the 3.6-m TNG telescope and its Sarg spectrograph
\citep{Gratton:01::}. Since fall 2006, the TATOOINE project has been collecting velocity 
measurements at Lick Observatory using the 3-m Shane and 0.6-m Coude Auxiliary 
Telescope (CAT) telescopes and the Hamilton Spectrograph \citep[Hamspec;]{Vogt:87::}. 
Overall, $\sim 50$ northern binaries are currently being monitored by the TATOOINE project. 
Recently, we have also initiated a southern effort at the 3.9-m AAT telescope with 
the UCLES spectrograph \citep{Diego:90::}.

The spectrographs employed in the survey, Hires, Sarg and Hamspec, are all
high resolution echelle spectrographs equipped with iodine absorption cells. 
They provide spectra with a resolution of 67 000, 86 000 and 60,000 respectively 
across a $\sim$400-900 nm bandwidth centered on the 500-600 nm range corresponding 
to an iodine cell's spectral features. The cells are used to superimpose a reference
absorption spectrum in a manner described in \S{2}. For each measurement, 
back-to-back spectra of the target binary were acquired, first with an 
iodine absorption cell in the path of the starlight, then with the iodine 
cell removed. Depending on the telescope, different exposure times were
used to balance the resulting SNR per collapsed pixel and the efficient 
use of telescope's time. The highest SNR of $\sim$250 
was achieved with the Keck I/Hires also to test the data pipeline in the regime 
of high SNR spectra. The typical SNR was $\sim$75-150 for the TNG/Sarg and $\sim$50-150 
for the Shane/CAT/Hamspec. In consequence, the RVs have different precision
ranging from $\sim$2 (HD195987, the best case) to 20 m$\,$s$^{-1}$ for the primary 
stars. In any case, the precision is sufficient to detect planets with masses 
as small as 0.3 M$_{Jup}$. Note also that due to a brightness ratio between the
primary and secondary, the RVs of the secondary are typically of lower
precision as the SNRs are for the composite observed spectra. For example,
an SNR of 250 and a brightness ratio of 6.7 (HD195987) corresponds to an
SNR of 220 for the primary and only 30 for the secondary.

A circumbinary planet will exhibit two indirect effects on the RV of the
stars.  First, the apparent system velocity will vary in a periodic manner
due to the motion of the binary about the system barycenter, with amplitude
\citep{Mut:07::}
\begin{equation}
\Delta v_b  = 57\,\mbox{\rm m\,s}^{-1} \times 
\frac{ \left( M_p/M_{Jup} \right) \sin i_p}{
\sqrt{\left( \left( M_b+M_p
\right)/M_{\odot}\right)\left(a_p/1\mbox{\rm{AU}} \right)}}. 
\end{equation}
Differential reflex motion and perturbations of the binary orbit by the
planetary companion are expected to be negligible on reasonable timescales.
Second, the finite speed of light will cause apparent changes in the phase
of the binary orbit due to the reflex motion of the binary about the
system center of mass.  This phase shift is detected for planets with
masses as small as \citep{Mut:07::}
\begin{equation}
M_p = 70 M_{Jup} \times \frac{\left( \sigma_{rv}/{\rm 20\,m/s}\right) 
\left(P_b/{\rm 5\,d}\right)^{4/3} \left(M_b/M_{\odot}\right)^{2/3}}
{\sqrt{N-6}\sin{i_b}\sin{i_p}\left( a_p/{\rm 1\,AU}\right)}
\end{equation}
\noindent assuming $1\%$ FAP (SNR=5.8) where $N$ is the number of observations and $i_b$ 
and $i_p$ are the inclinations of the binary and planet orbits, respectively. Note that 
this observable is actually {\em more} sensitive to longer period planets.  
In combination, these effects allow us to detect Jupiter mass planets or 
smaller for an extended range of orbital periods. In the current analysis,
the second effect has been ignored as it is small on the relatively short 
timescales being considered.

\section{Planet detection limits for ten SB2s}

For this initial sample of stars, we have selected these SB2s from our survey
that have long time span RV data sets and an orbital phase coverage allowing 
for a reliable tomographic disentangling. These are HD 9939, HD 13974,
HD 47415, HD 78418, HD 86146, HD 195987, HD 210027, HD 214686, HD 221950 and
HD 282975. Let us note that we recently made improvements to our data pipeline 
and the Keck I/Hires RVs used in this paper are typically several m$\,$s$^{-1}$ 
more accurate than those shown  in \cite{Konacki:09::}. The internal RV
errors are computed from the scatter between the echelle orders used in the
reduction. They are expected to underestimate the real RV scatter. In
addition to the stellar jitter, one of the main reasons for the
underestimation of the errors are the imperfections in the disentangled 
template spectra. While these imperfections are impossible to see with 
a``naked  eye'', they still contribute to the total error budget. For this reason 
we add an additional error in quadrature to obtain a reduced $\chi^2$ equal to 1
for a simple Keplerian model. The RVs are first modeled with a Keplerian RV model 
and a least-squares orbital fit is made simultaneously to the velocities of the 
primaries and secondaries (the orbital solutions and RVs will be published separately). 
The residuals are then inspected for planetary signatures which would obviously be 
the same for a primary and secondary. In the process, we have not found any clear 
planetary signatures. 

The procedure of \cite{Cum:99::} for evaluating the regions in
Mass-Period space in which companions can be ruled out was modified for
application to the fit residuals to the 2-body Keplerian orbits of the
target binaries in order to search for and place limits on additional 
components as follows. This implementation of the analysis algorithm 
was thoroughly tested during the SIM Double-Blind-Test
\citep{Traub2009a,Traub2009b}.

A grid of potential companion orbital periods is sampled logarithmically at
values of $P = 2fT/I$, where $f$ is an optional excess factor (here 4) for
finer sampling, $T$ is the timespan of the observations, and $I$ is an integer 
beginning at 1 and continuing through that at which the sampled period is 
$\sim 5 \times$ the binary orbital period (at which point many companion orbits 
are dynamically unstable). The exact limiting orbital periods shown for each star
below were calculated using the equation (3) from \cite{Holman:99::}. At each 
sampled orbital period, the RV residuals for each star are fit to a Keplerian 
orbit with floating offsets as:
\begin{equation}
v = A\cos 2\pi t/P+B\sin 2\pi t/P+v_{\circ, i}
\end{equation}
where $v_{\circ, i}$ is a floating velocity offset (the $i$ representing that
different floating offsets are used for each star and each observatory;
e.g.~for binary stars with Keck data from before and after the detector upgrade,
TNG-SARG data, and Lick data, a total of 8 different velocity offset
parameters are used). At each sample period, the $\chi^2$-minimizing 
values of $A$, $B$, and $v_{\circ, i}$ and resulting $\chi^2$ value are 
evaluated. The circular orbit model is found to be sensitive to mildly 
eccentric companions, as in \cite{Cum:99::}.

The $\chi^2$ values of the best-fit circular orbits were recorded for 
each potential companion period. These were converted to the $z$-statistic 
of \cite{Cum:99::} for plotting signal power in a periodogram as:
\begin{eqnarray} z &=& Q \frac{\chi_\circ^2 - 
\chi^2}{\chi^2_{\rm min,\, circ}}\\
Q_{\rm circ} &=& \frac{ N_D - (2+N_{v_{\circ, i}})}{2}
\end{eqnarray}
where $N_D$ is twice the number of double-RV measurements, $N_{v_{\circ,
i}}$ is the number of independent velocity offsets $v_{\circ, i}$, 
$\chi_\circ^2$ is the fit to no orbit at all ($v = v_{\circ, i}$). (When 
making comparisons for calculating false alarm probabilities and detection 
limits, the $Q$ coefficients cancel and their actual values have little impact.)

Once the maximum value of $z$ has been found for the entire data set, the
false alarm probability (FAP) of that largest signal is calculated as
follows. A Gaussian random number generator is used to create 10,000
synthetic data sets with no Keplerian signal but with the same cadence and
measurement uncertainties as the original data. This introduces a slight
difference with the procedure of \cite{Cum:99::}: a random number generator 
is used instead of scrambling and rescaling the fit residuals. As noted by those
authors, the difference in these approaches causes little change in the
results. These synthetic zero-signal data sets are analyzed with the same 
procedure as for the original data residuals. Because these synthetic data 
sets are known to contain no real signal, the fraction of these whose analysis 
show power larger than the maximum $z$ of the real data set determines how likely 
that most significant value is to be a false positive.

Finally, the threshold companion RV signal (as a function of orbital period)
that can be excluded using the current data residuals is evaluated. At each
sample period, an initial guess for the RV amplitude $K$ that can be
excluded is made, and 10,000 synthetic data sets with a signal of that
amplitude are generated. The orbital phase is selected randomly with flat
distribution. The fraction of orbits with $z$ exceeding
that of the data is computed, the RV amplitude $K$ is modified, and the
procedure iterated until the fraction is constrained near the desired
reliability value (here 99\%).  
In Figures 6-9 (right panels), the $z$ periodograms are shown for each binary, with a
horizontal line at the 1\% FAP level as determined from the signal-free 
synthetic data sets. In Figures 6-9 (left panels), the Mass(sin i)-Period phase-space 
in which circumbinary companions can be excluded at the 99\% confidence 
level are shown, where the values of Mass(sin i) are calculated from the
threshold companion RV amplitudes $K$ using the total masses of the binary,
as estimated for each system. We have also tested orbits with non-zero 
eccentricities. It turns out that moderate eccentricities make little 
impact as the threshold lines move up by a factor of about 1.5. This is
demonstrated on the case of HD9939 in Figure 10.

The targets and their RV data sets are summarized in Table 1 where 
Sp denotes the target's spectral type (either combined or for
each component), M$_{1,2}$ the masses of the primary and secondary used
in the analysis, P$_{orbital}$ the orbital period. a$_{stable}$ and P$_{stable}$
are the semi-major axis and orbital period of the first stable orbit of a 
circumbinary planet computed using the equation (3) from \cite{Holman:99::}. 
N$_{1,2}$ denotes the number of available RV measurements for the entire data 
set and each subset, rms$_{1,2}$ the corresponding best-fit rms from a Keplerian 
orbit for the entire data set as well as subsets, T$_{span}$ the time span of the 
entire RV data set. $\sigma_{1,2}$ are the formal RV errors and $\epsilon_{1,2}$ 
are the additional errors added in quadrature.

\section{Conclusions}

Our novel iodine cell based RV technique allows one to measure precise
RVs of the components of double-lined spectroscopic binary stars. With
this technique, in 2003 we have initiated TATOOINE, a radial velocity search 
for circumbinary planets around a sample of $\sim$50 SB2s. In this paper
we present the first results from this survey --- a non detection
of exoplanets in the 0.3-3 M$_{Jup}$ regime with the orbital periods
of up to 5.3 years around 10 SB2s. 

Recently two circumbinary planets around an eclipsing binary HW Vir 
\citep {Lee:09::} and a circumbinary brown dwarf around en eclipsing binary 
HS0705+6700 \citep{Q:09::} have been claimed to be detected by means of eclipse 
timing. This is however not the first time when substellar companions or planets
have been detected with a timing technique. In addition to the confirmed
case of the three rocky planets around a millisecond pulsar B1257+12
\citep{Konacki:03::}, planets have been claimed to orbit a pulsar B0329+54
based on the timing of its radio pulses \citep{Dem:79::,Shab:95::}.
Later it was demonstrated that the timing variation is quasi-periodic
and is not due to planets \citep{Konacki:99::}. One is left to wonder if 
the two cases of the eclipse timing variations are indeed best explained 
by a periodic signal due to circumbinary bodies and not an unrecognized 
quasi-periodic phenomenon mimicking a periodic planetary signal.

Our sample is too small to risk any decisive conclusion about the frequency 
of circumbinary planets. This reminds us of the pioneering search for planets 
around single stars by \cite{Campbell:88::}. In particular, their
non-detection of the population of hot Jupiters. We also have not detected
any short period planets or to be precise planets with periods near the
inner orbital stability limits. The question of the existence of such 
circumbinary planets and hence the impact of the central body's binarity 
on the parking mechanisms for migrating planets remains open. It should
also be noted that even though all our targets are SB2s, they constitute
quite a diverse sample. Their orbital periods range from 5.7 to 57 days
and the primary to secondary mass ratios range from 1 to 1.6. This
presumably should make an impact on the formation and evolution of
circumbinary planets. The targets from RV surveys for planets around single 
stars are more homogenous and provide for an overall similar environment.

When comparing our planet detection limits to those from the Anglo-Australian 
Planet Search \citep{OT:09::}, the McDonlad Obseravtory Planet Search \citep{Wit:06::} 
and the Keck Planet Search \citep{Cum:08::} (all employing iodine cells), one 
will notice that these surveys allow for a detection of up to several times less 
massive planets. This is due to a higher precision (now 
approaching 1 m$\,$s$^{-1}$ in many cases), typically larger number of RV
measurements and a longer time span of the data sets. On the other hand, 
the comparison with the planet detection limits for the Lick Planet Search 
\citep{Cum:99::}, the first RV survey to employ an iodine cell, demonstrates
that our survey performs similarly. One should also remember that the mass of 
the central body in our survey is typically 1.5-2.0 times higher then in the 
case of single stars which obviously decreases the sensitivity to planets 
in terms of masses by such a factor.

Our search continues and the changes to the method are constantly being
made to improve the RV precision. Theoretical works on the formation and 
evolution of circumbinary planets are strongly encouraged as now we have 
the observational tools to test them.

\acknowledgments

We thank Lucasfilm Ltd for inspiring the TATOOINE planet search (and careers
of many of us), and Lucasfilm's Senior Director of Business Affairs 
David Anderman for an excellent tour of the Lucasfilm complex upon hearing
about our program.  The tour of Lucasfilm was a highlight of the 
undergraduate research experiences of Agnieszka Czeszumska, Sam Halverson, Tony 
Mercer, and Jackie Schwehr. We thank the California and Carnegie Exoplanet Search team,
and Geoff Marcy in particular, for allowing us access to their precision
velocimetry tools at Lick Observatory. This research has made use of the Simbad 
database, operated at CDS, Strasbourg, France.  MWM acknowledges support from the 
Townes Fellowship Program. M.K. is supported by the Foundation for Polish Science 
through a FOCUS grant and fellowship, by the Polish Ministry of Science and Higher 
Education through grants N203 005 32/0449 and 1P03D-021-29. Part of the
algorithms used in this analysis were developed during the SIM Double Blind 
Test, under JPL contract 1336910.  This research has made use of the Simbad 
database, operated at CDS, Strasbourg, France. The observations on the
TNG/SARG have been funded by the Optical Infrared Coordination network (OPTICON),
a major international collaboration supported by the Research Infrastructures 
Programme of the European Commissions Sixth Framework Programme.

{\it Facilities:} \facility{Keck I/Hires, TNG/Sarg, Shane/Hamspec}

\clearpage

\begin{figure}
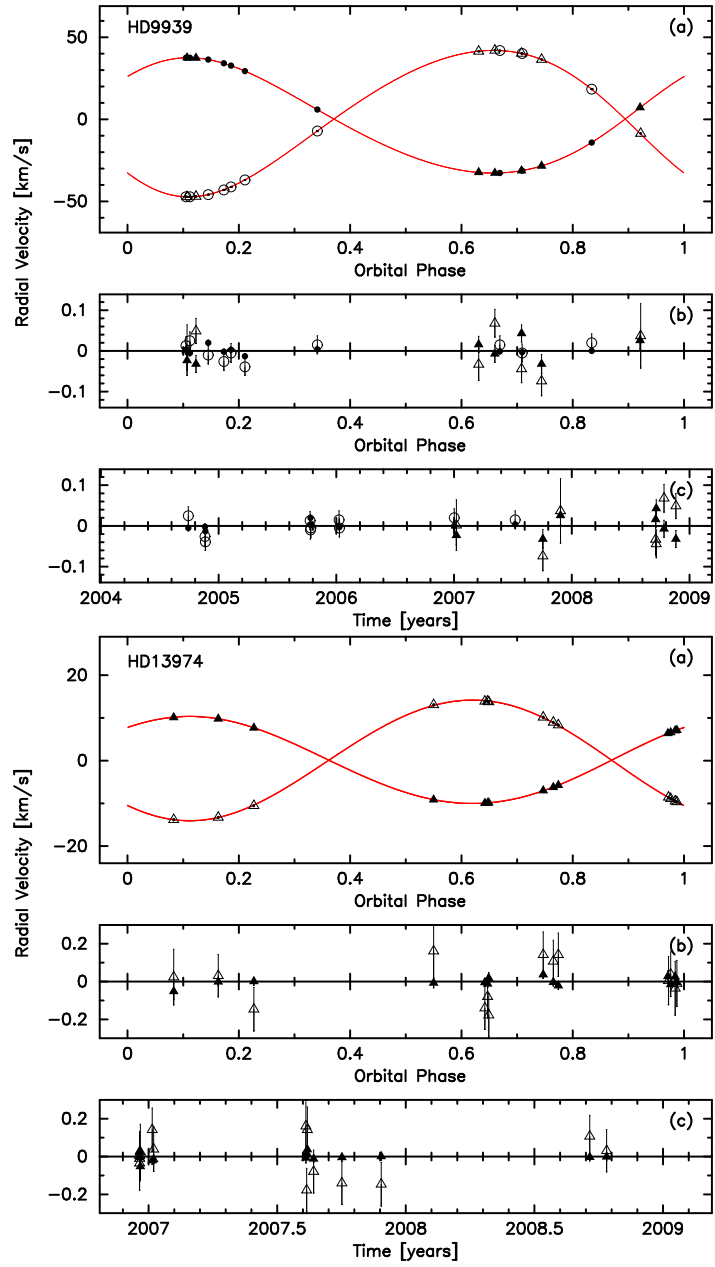

\epsscale{0.61}
\plotone{fig1a.ps}

\plotone{fig1b.ps}
\caption{The RVs of HD9939 (top) and HD13974 (bottom) as a function of the orbital phase
(a), and the residuals (observed minus modeled RVs) as a function of the orbital phase 
(b) and time (c). The primary is denoted with filled symbols, the secondary with open 
ones and the best-fit RV model with a solid line. The Keck I/Hires is denoted with 
circles, Shane/CAT/Hamspec with triangles and TNG/Sarg with stars. 
\label{fig1}}
\end{figure}

\begin{figure}
\epsscale{0.61}
\plotone{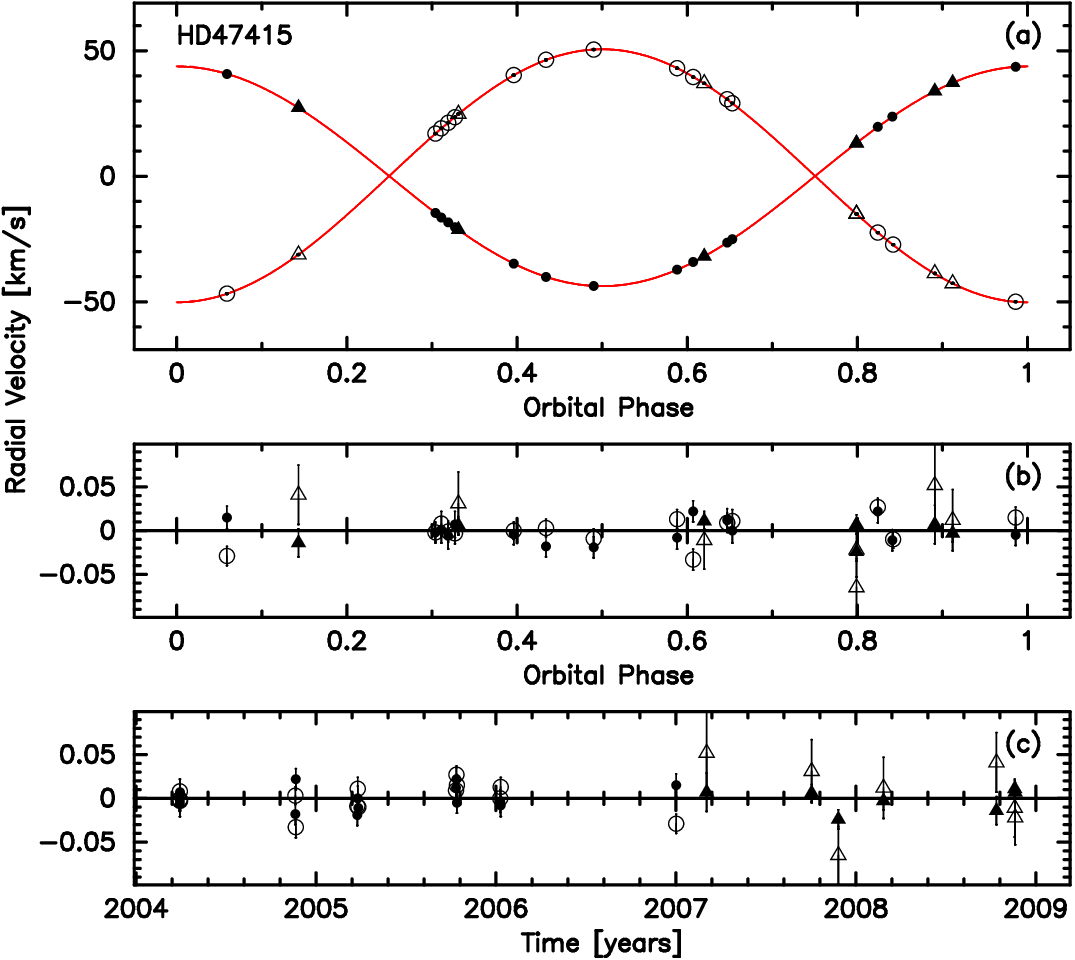}

\plotone{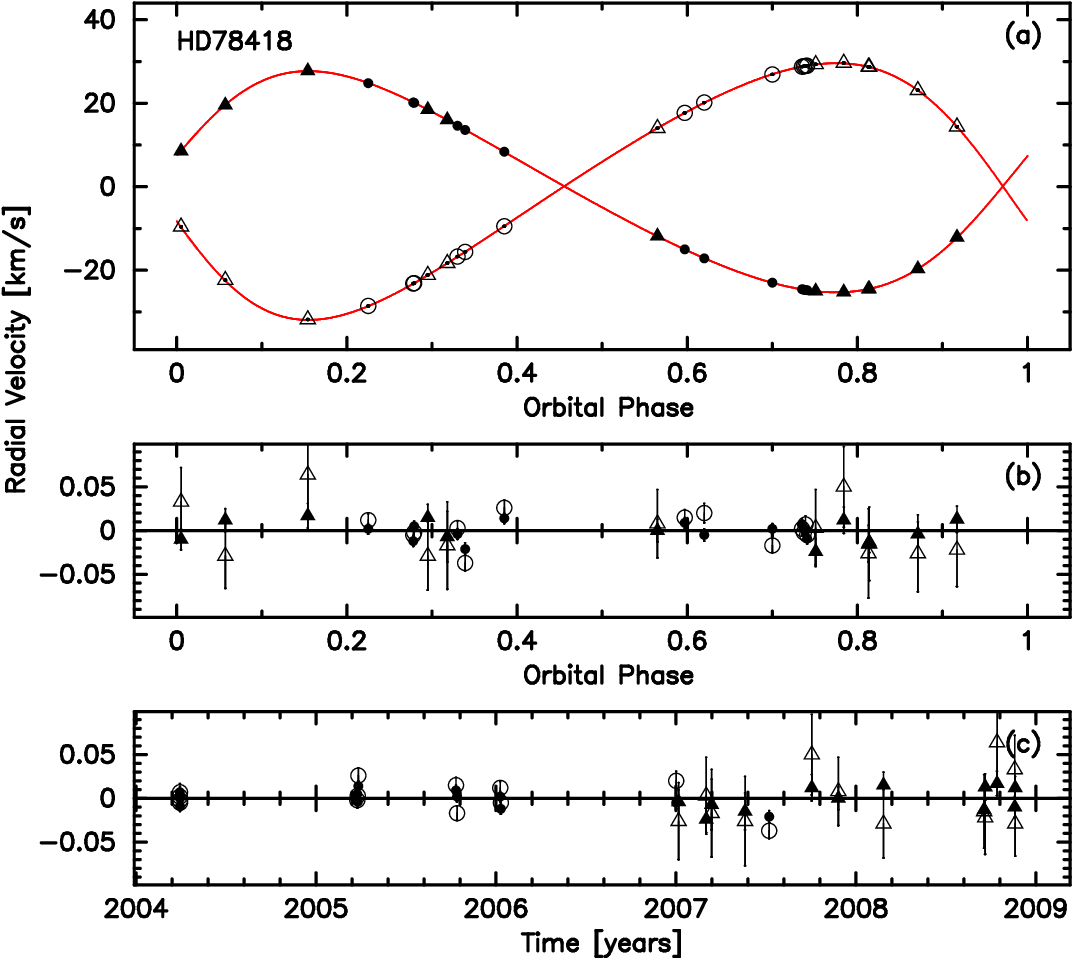}
\caption{The RVs of HD47415 (top) and HD78418 (bottom) as a function of the orbital phase
(a), and the residuals (observed minus modeled RVs) as a function of the orbital phase 
(b) and time (c). The primary is denoted with filled symbols, the secondary with open 
ones and the best-fit RV model with a solid line. The Keck I/Hires is denoted with 
circles, Shane/CAT/Hamspec with triangles and TNG/Sarg with stars.
\label{fig2}}
\end{figure}

\begin{figure}
\epsscale{0.61}
\plotone{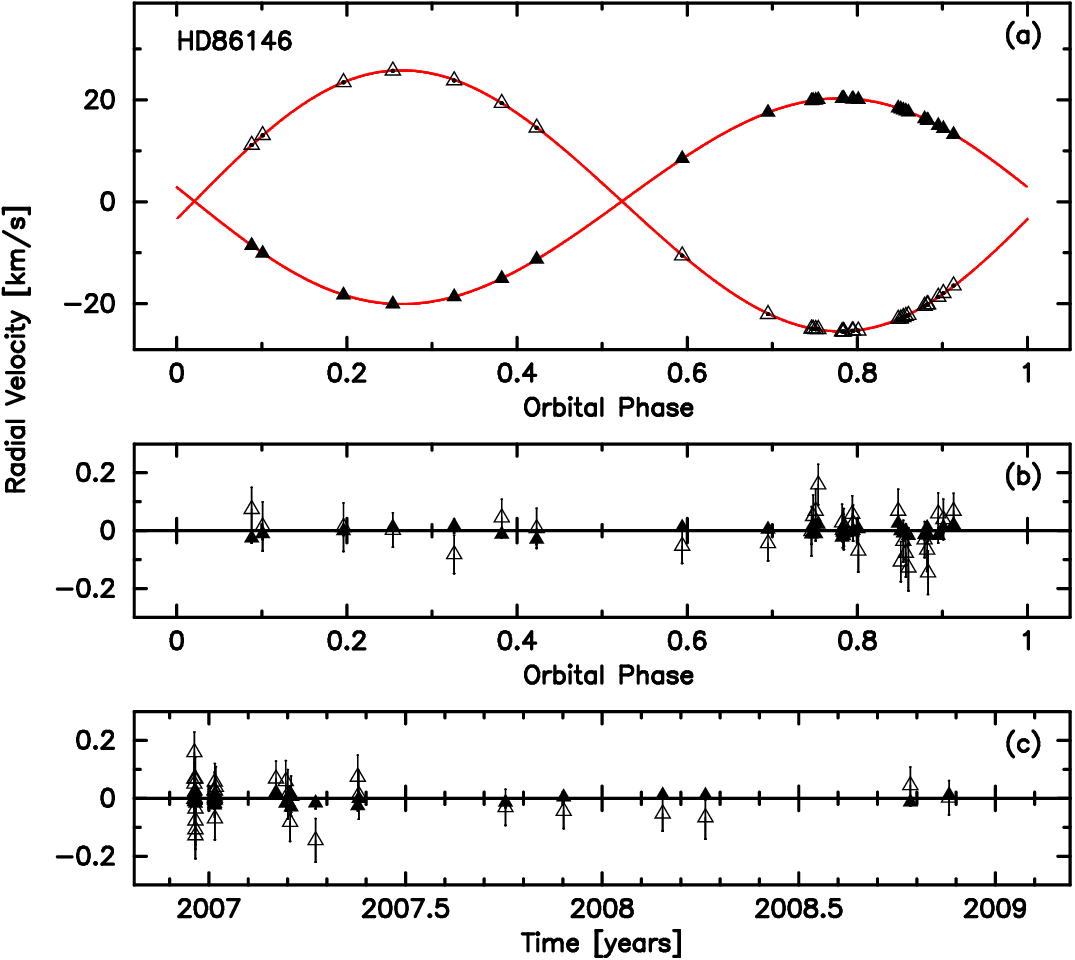}

\plotone{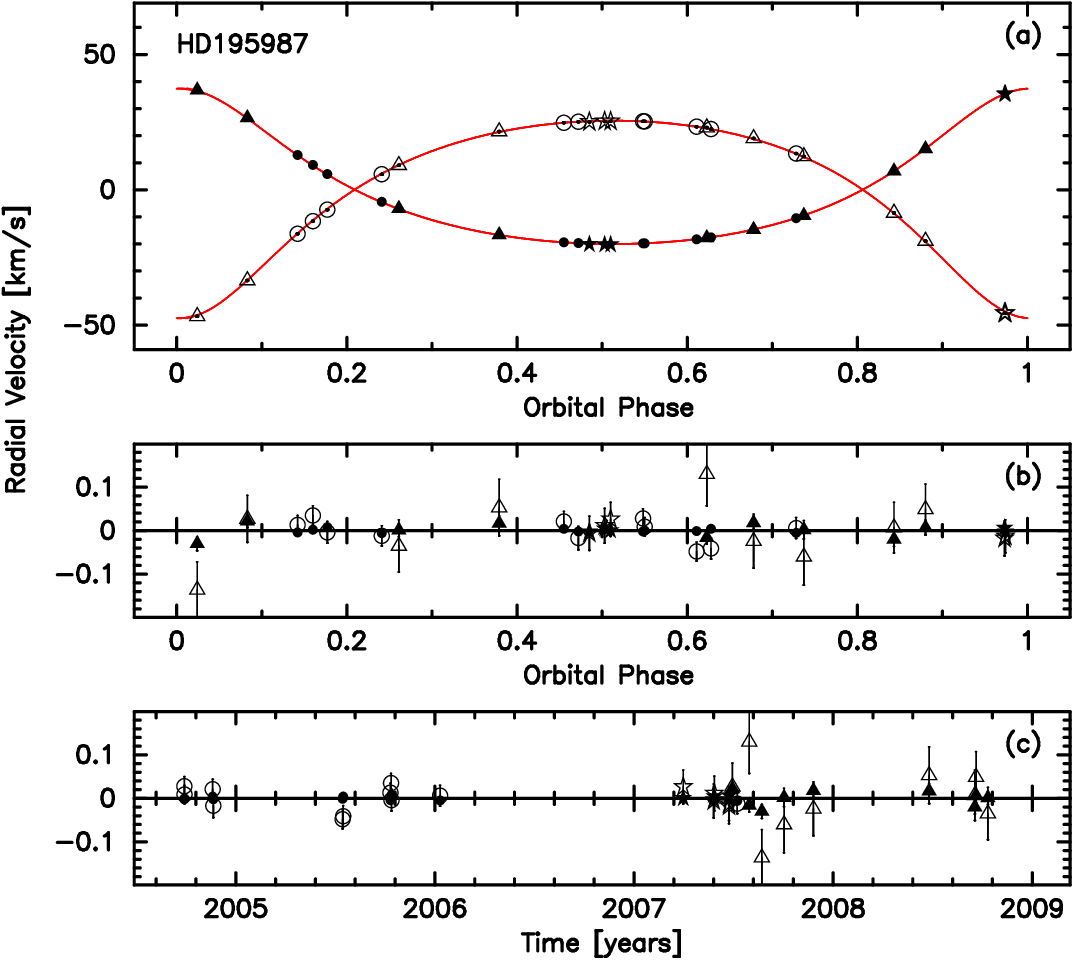}
\caption{The RVs of HD86146 (top) and HD195987 (bottom) as a function of the orbital phase
(a), and the residuals (observed minus modeled RVs) as a function of the orbital phase 
(b) and time (c). The primary is denoted with filled symbols, the secondary with open 
ones and the best-fit RV model with a solid line. The Keck I/Hires is denoted with 
circles, Shane/CAT/Hamspec with triangles and TNG/Sarg with stars.
\label{fig3}}
\end{figure}

\begin{figure}
\epsscale{0.61}
\plotone{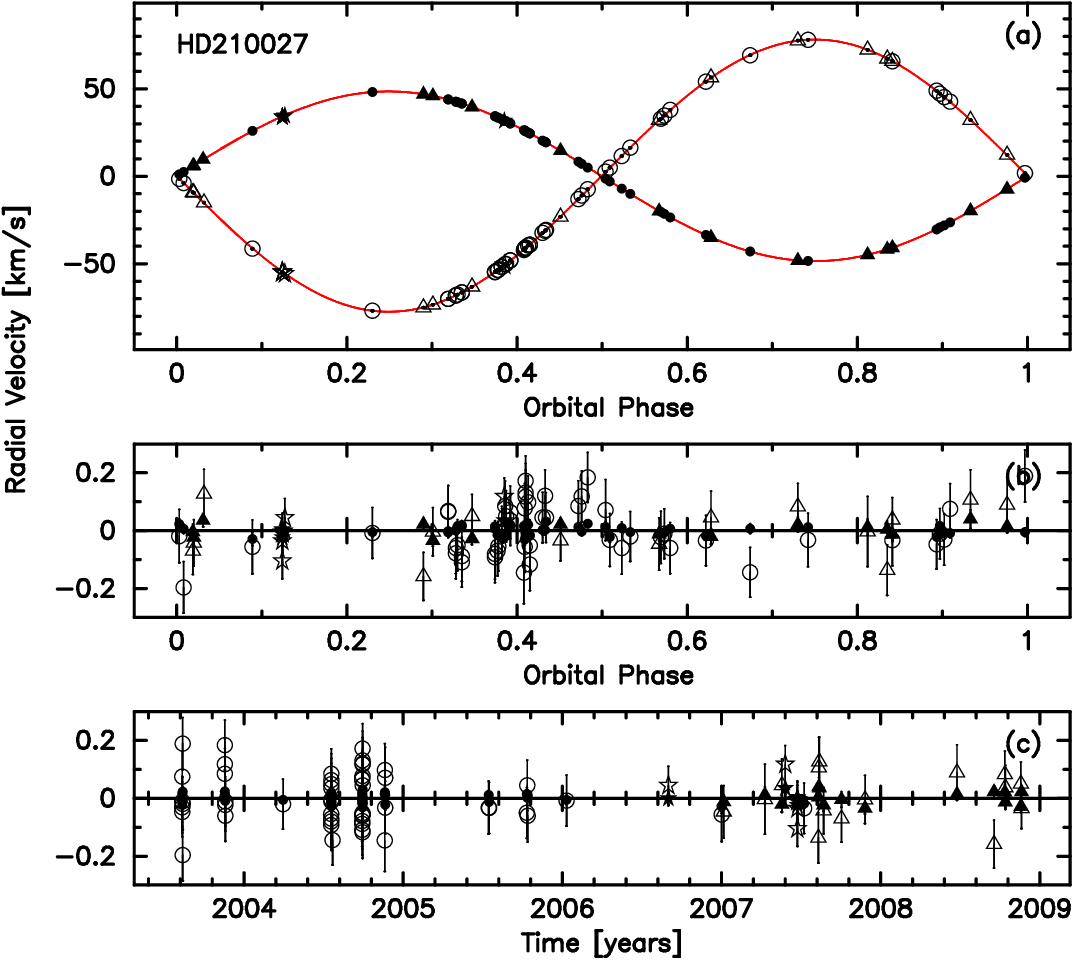}

\plotone{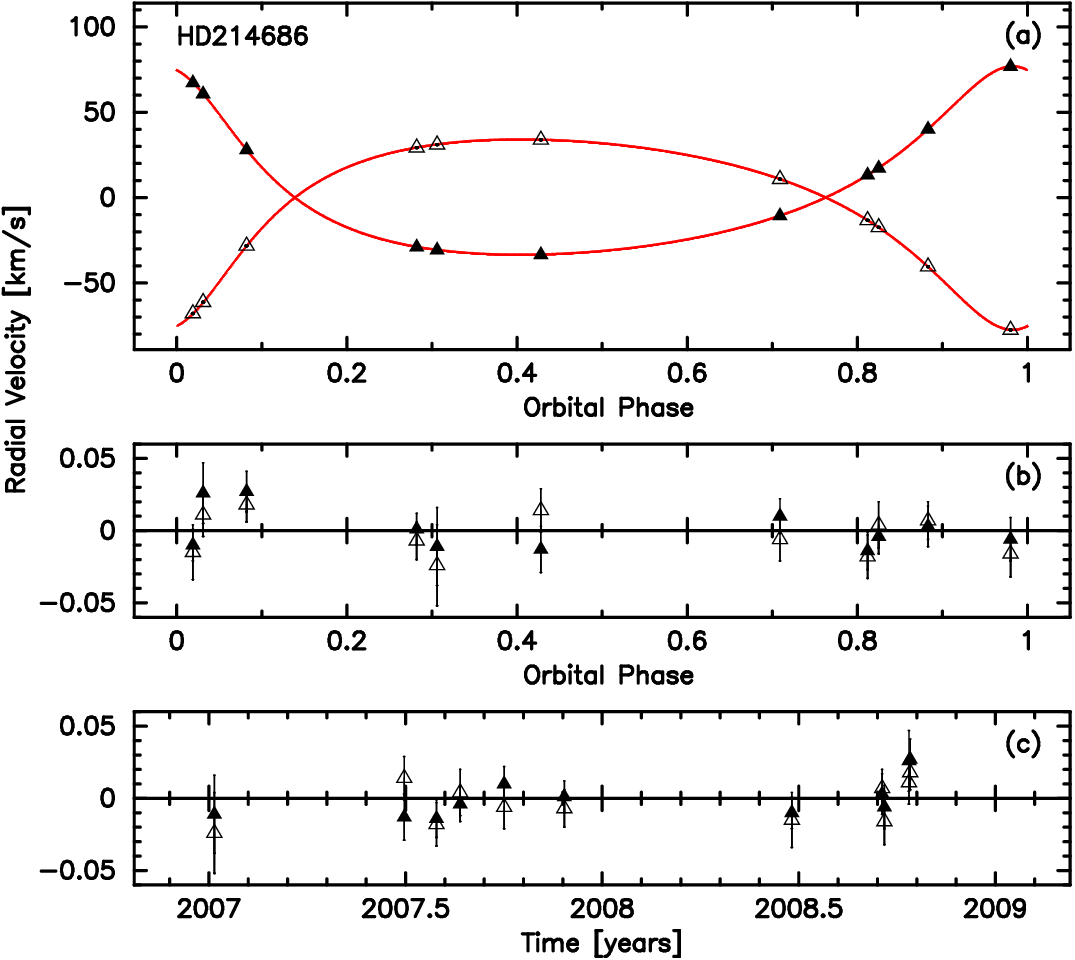}
\caption{The RVs of HD210027 (top) and HD214686 (bottom) as a function of the orbital phase
(a), and the residuals (observed minus modeled RVs) as a function of the orbital phase 
(b) and time (c). The primary is denoted with filled symbols, the secondary with open 
ones and the best-fit RV model with a solid line. The Keck I/Hires is denoted with 
circles, Shane/CAT/Hamspec with triangles and TNG/Sarg with stars.
\label{fig4}}
\end{figure}

\begin{figure}
\epsscale{0.61}
\plotone{fig5a.ps}

\plotone{fig5b.ps}
\caption{The RVs of HD221950 (top) and HD282975 (bottom) as a function of the orbital phase
(a), and the residuals (observed minus modeled RVs) as a function of the orbital phase 
(b) and time (c). The primary is denoted with filled symbols, the secondary with open 
ones and the best-fit RV model with a solid line. The Keck I/Hires is
denoted with circles, Shane/CAT/Hamspec with triangles and TNG/Sarg with stars.
\label{fig5}}
\end{figure}

\begin{figure}
\epsscale{0.85}
\plottwo{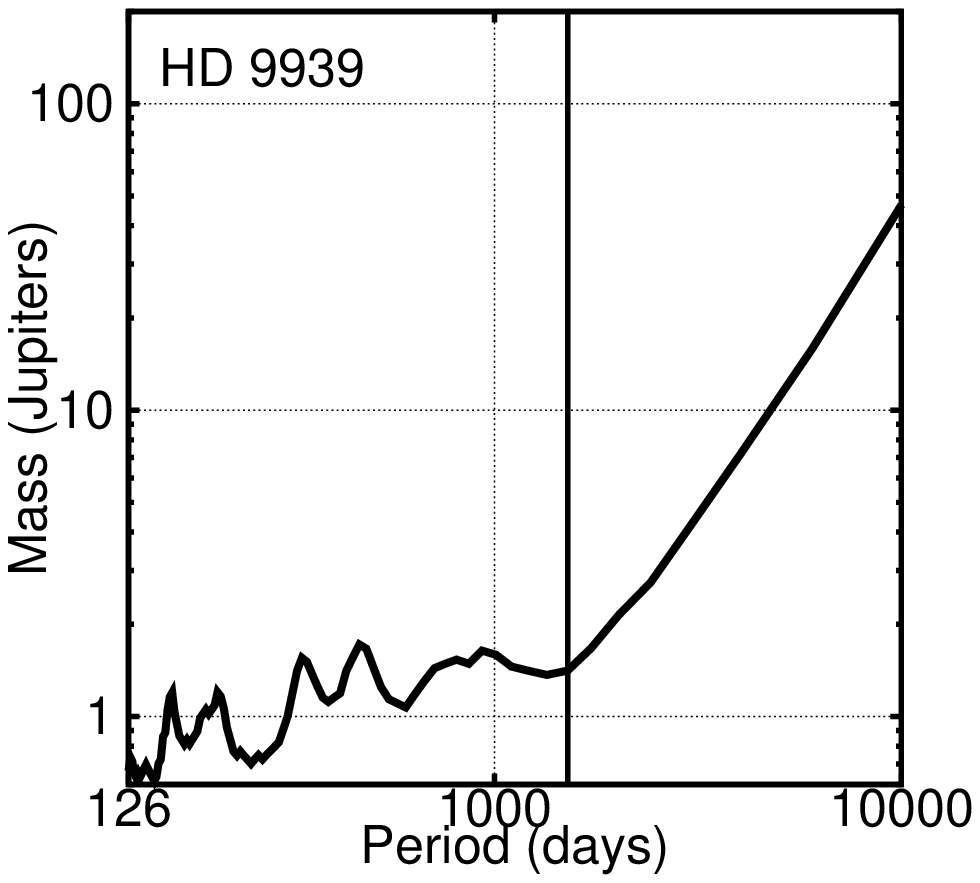}{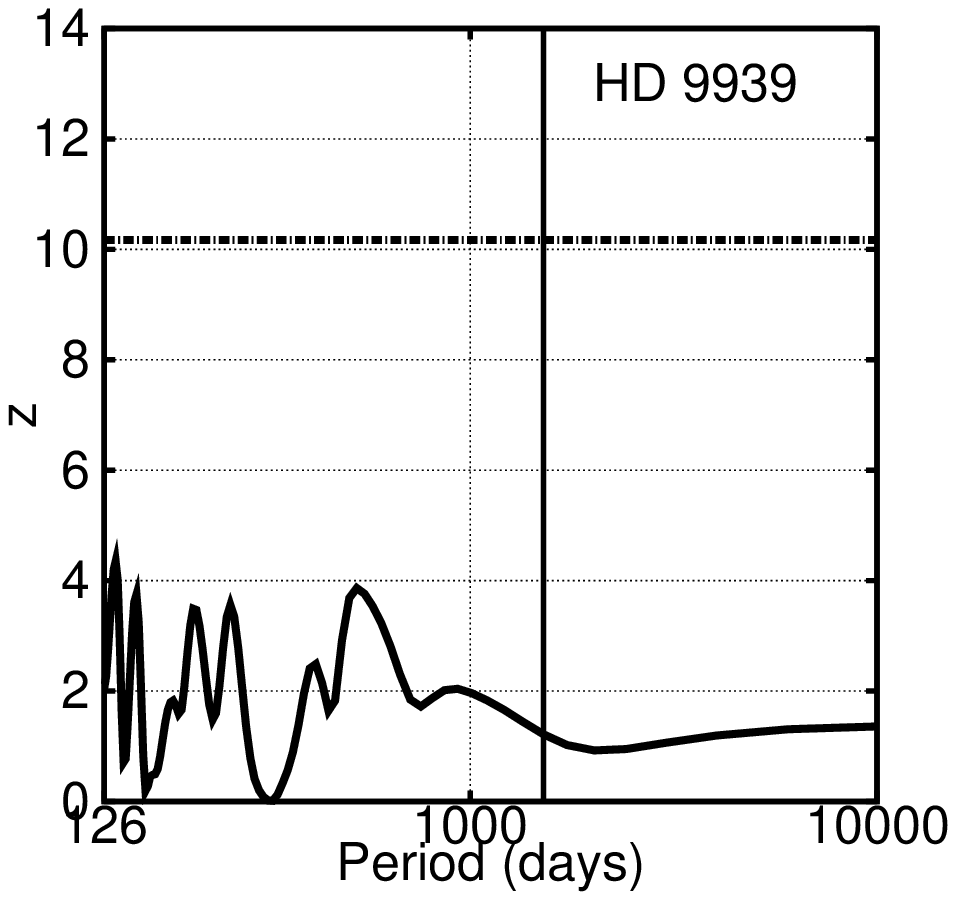}

\plottwo{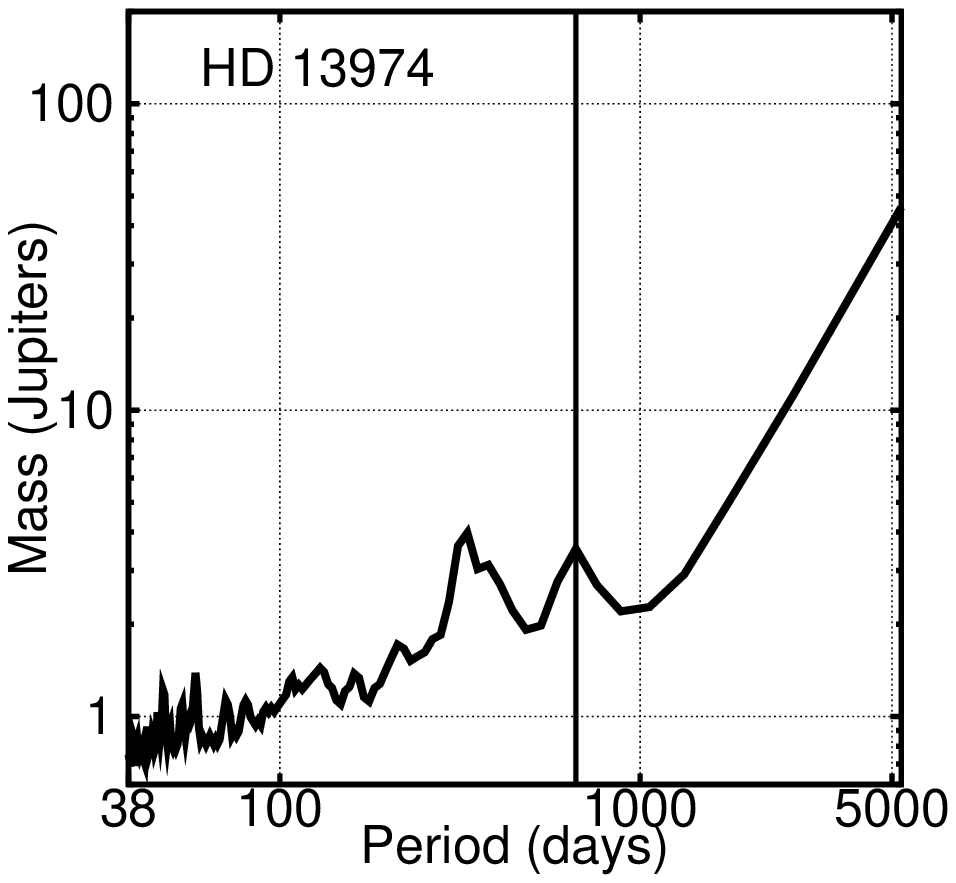}{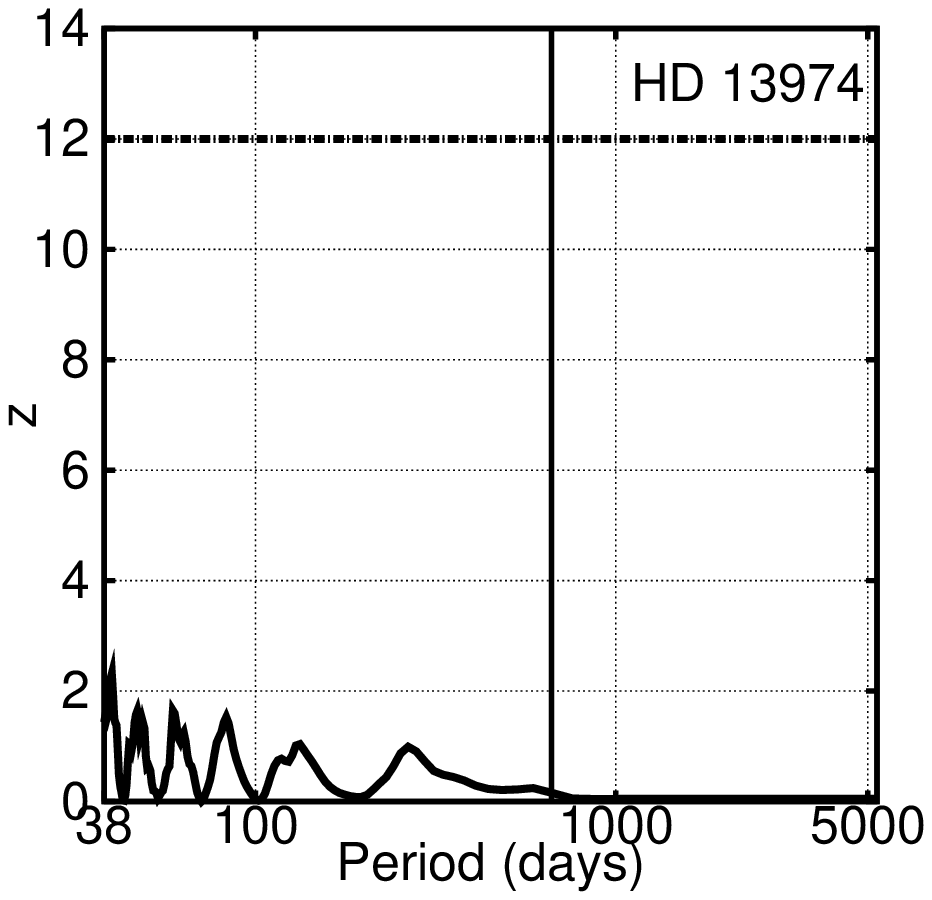}

\plottwo{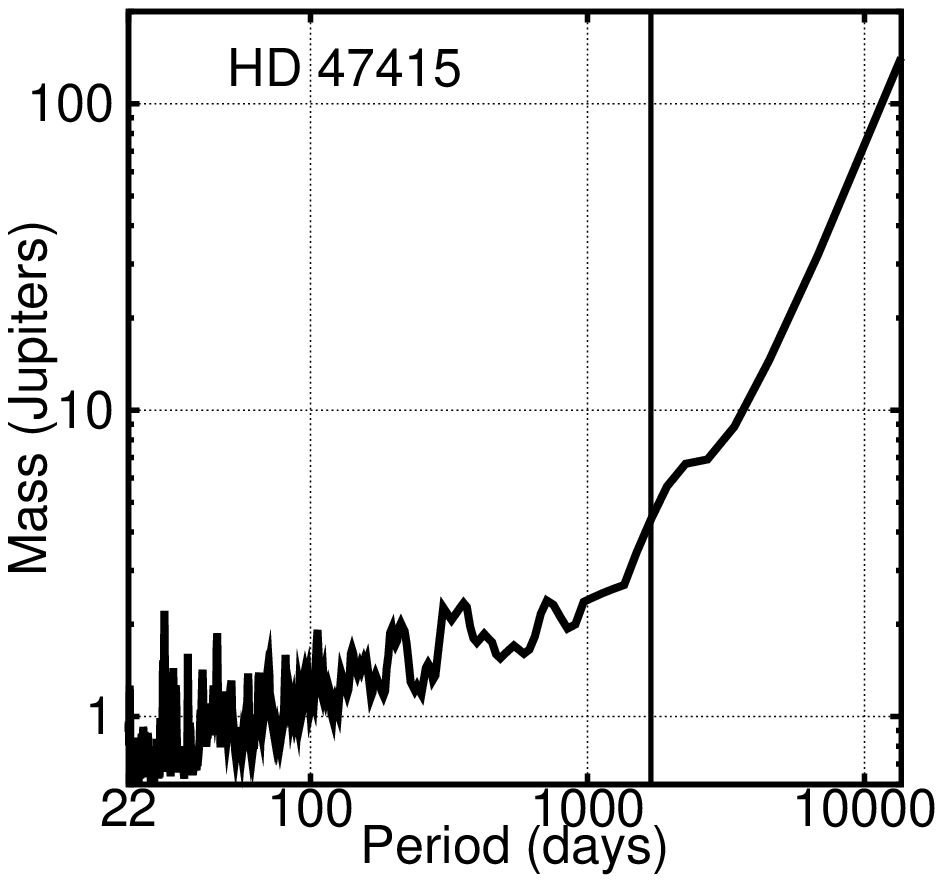}{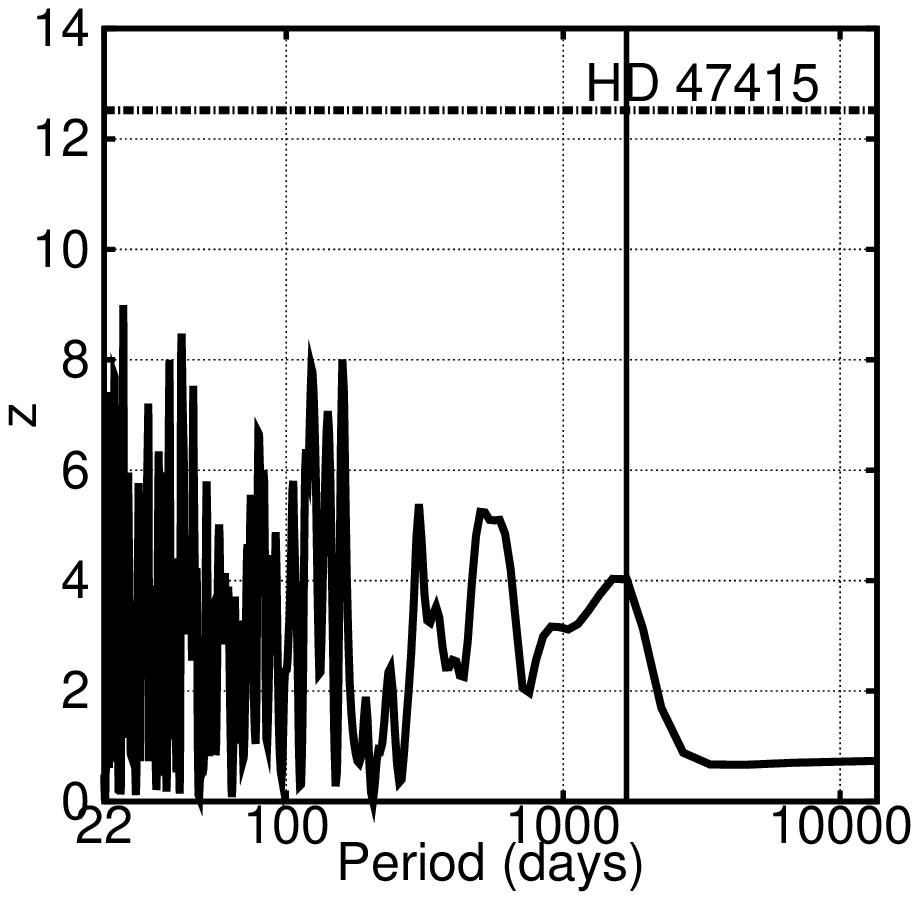}
\caption{Planet detection limits in the log P - log M space (orbital period
-mass; left panels) and periodograms (right panels). The solid line in the 
left panels is a planet detection limit corresponding to the 99\% confidence 
level. The vertical line near the orbital period of 1000 days denotes the
time span of the data set.
\label{fig6}}
\end{figure}

\begin{figure}
\epsscale{0.85}
\plottwo{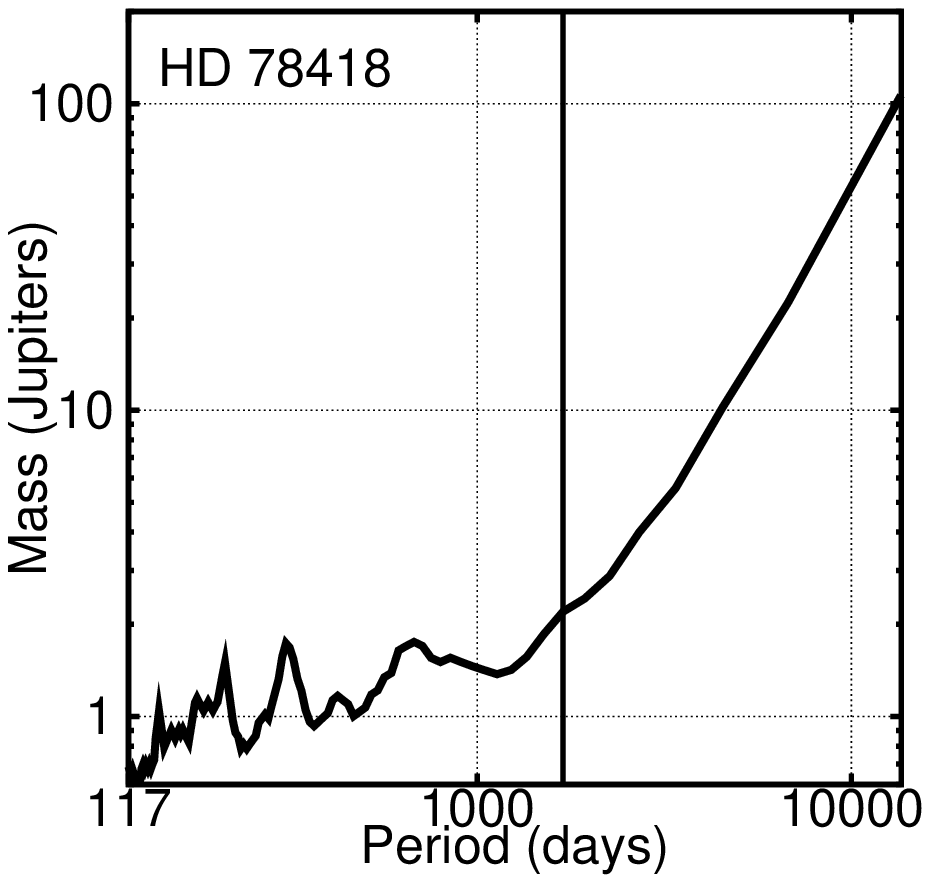}{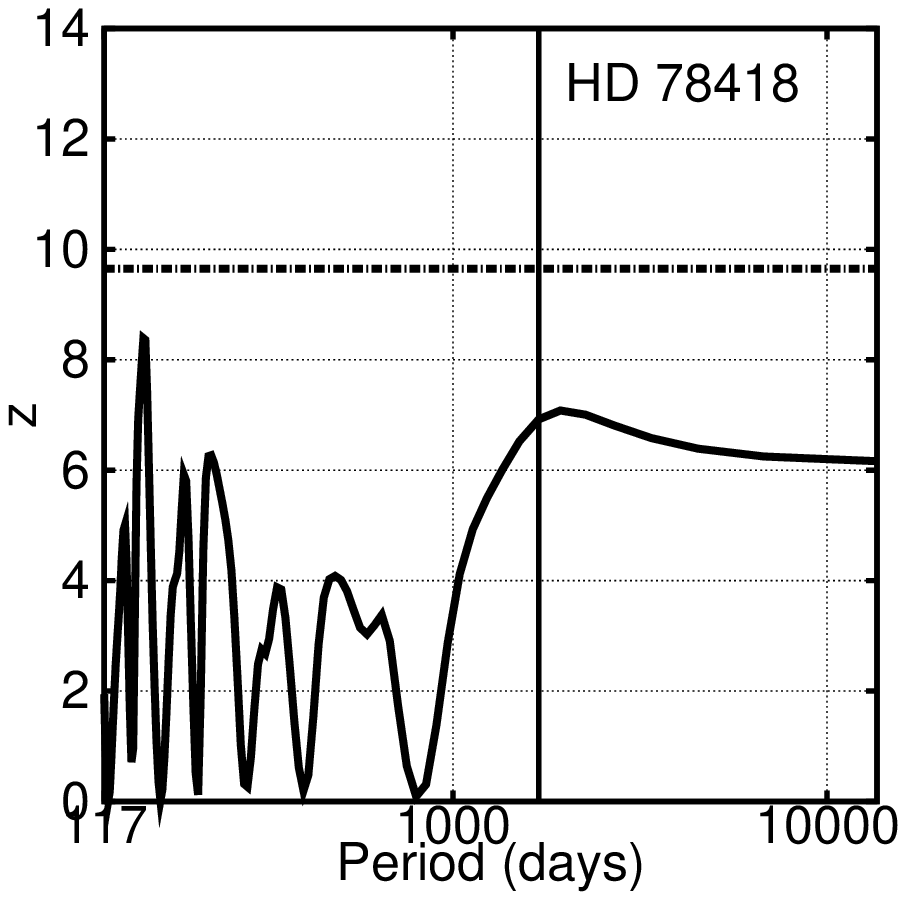}

\plottwo{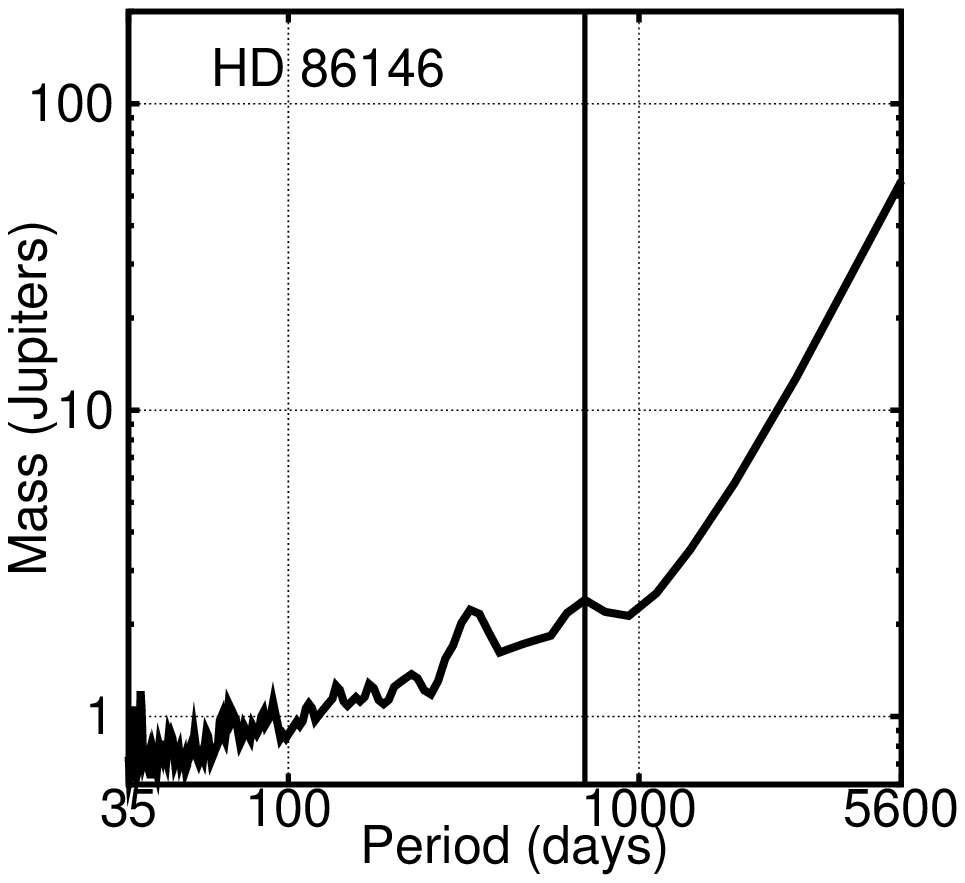}{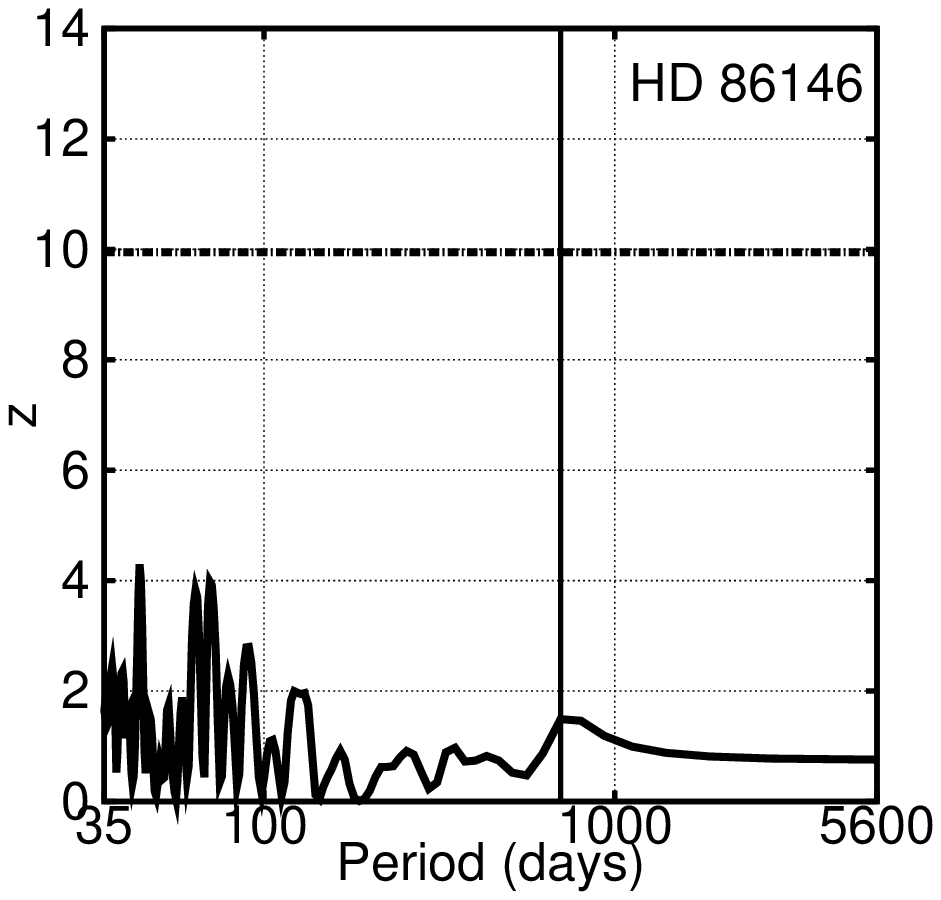}

\plottwo{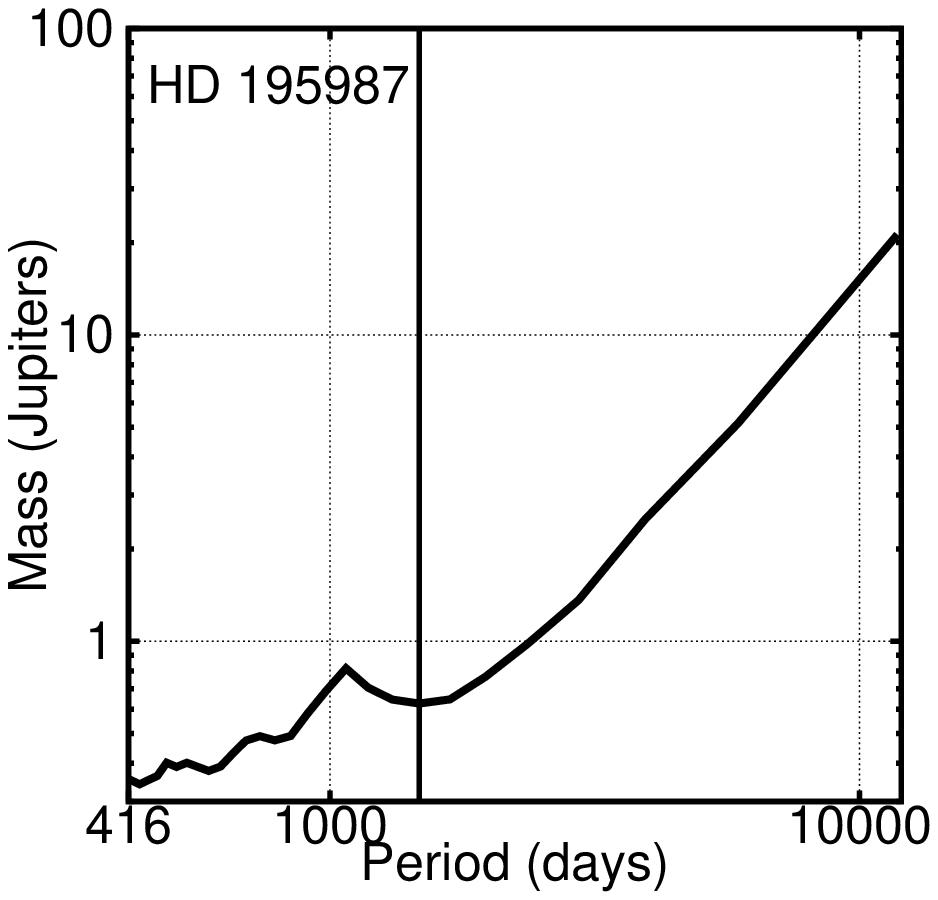}{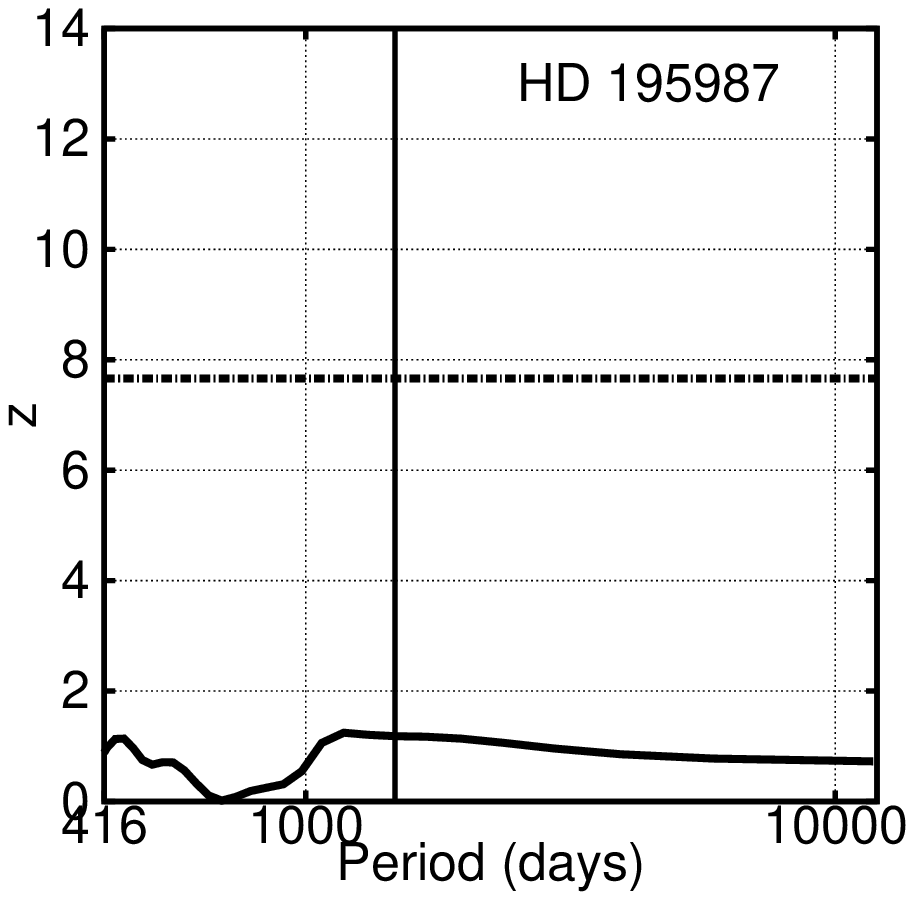}
\caption{Planet detection limits in the log P - log M space (orbital period
-mass; left panels) and periodograms (right panels). The solid line in the 
left panels is a planet detection limit corresponding to the 99\% confidence 
level. The vertical line near the orbital period of 1000 days denotes the
time span of the data set.
\label{fig7}}
\end{figure}

\begin{figure}
\epsscale{0.85}
\plottwo{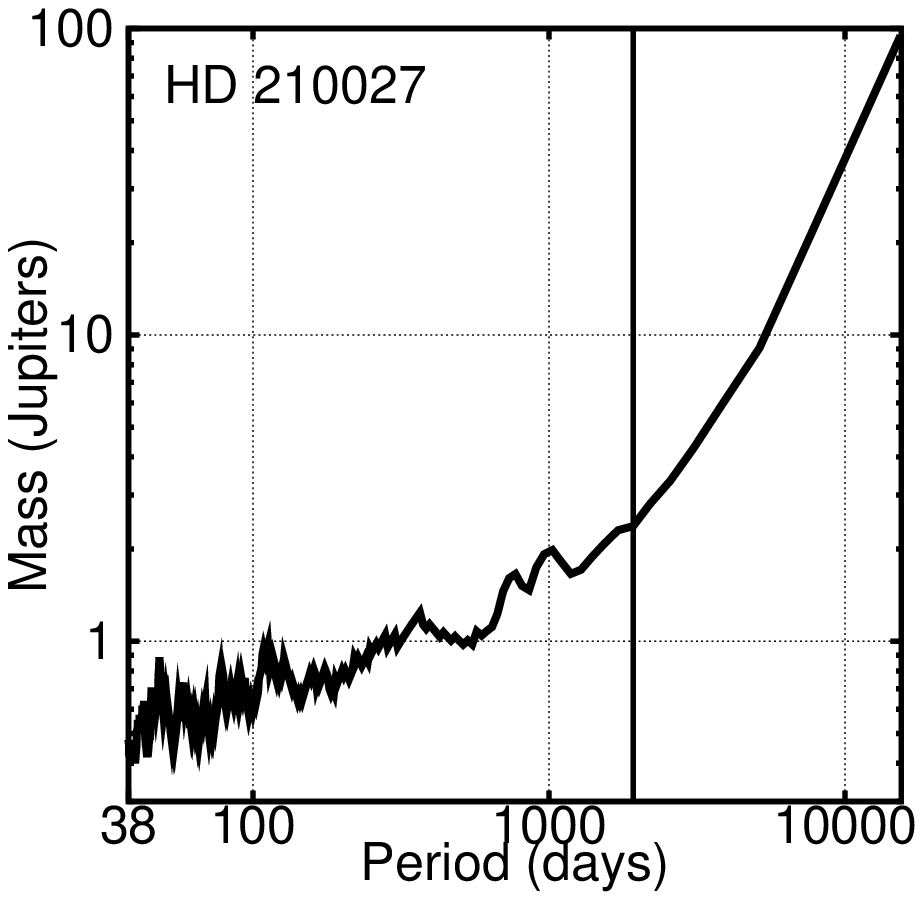}{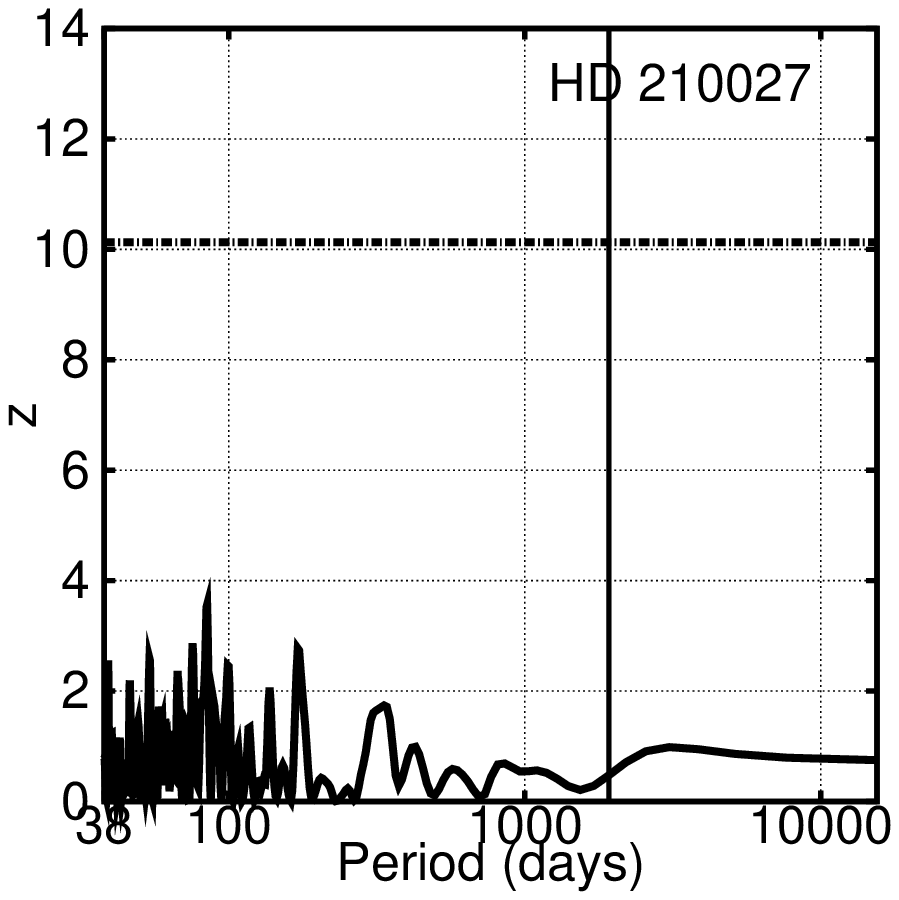}

\plottwo{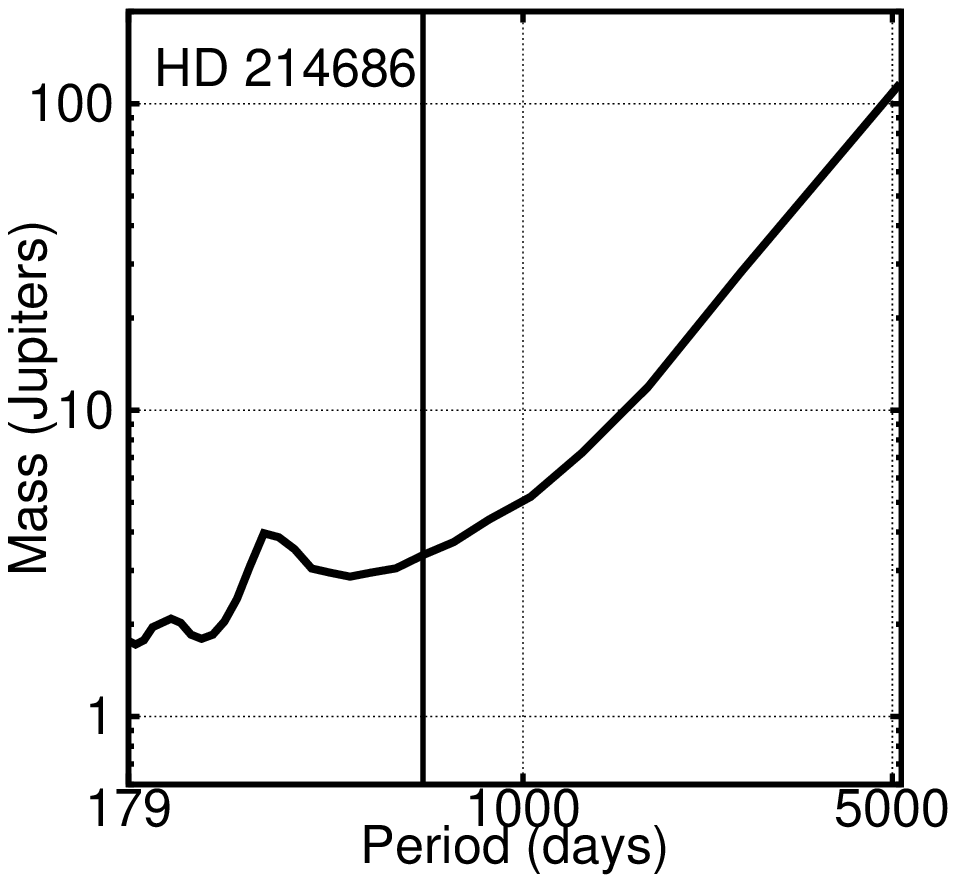}{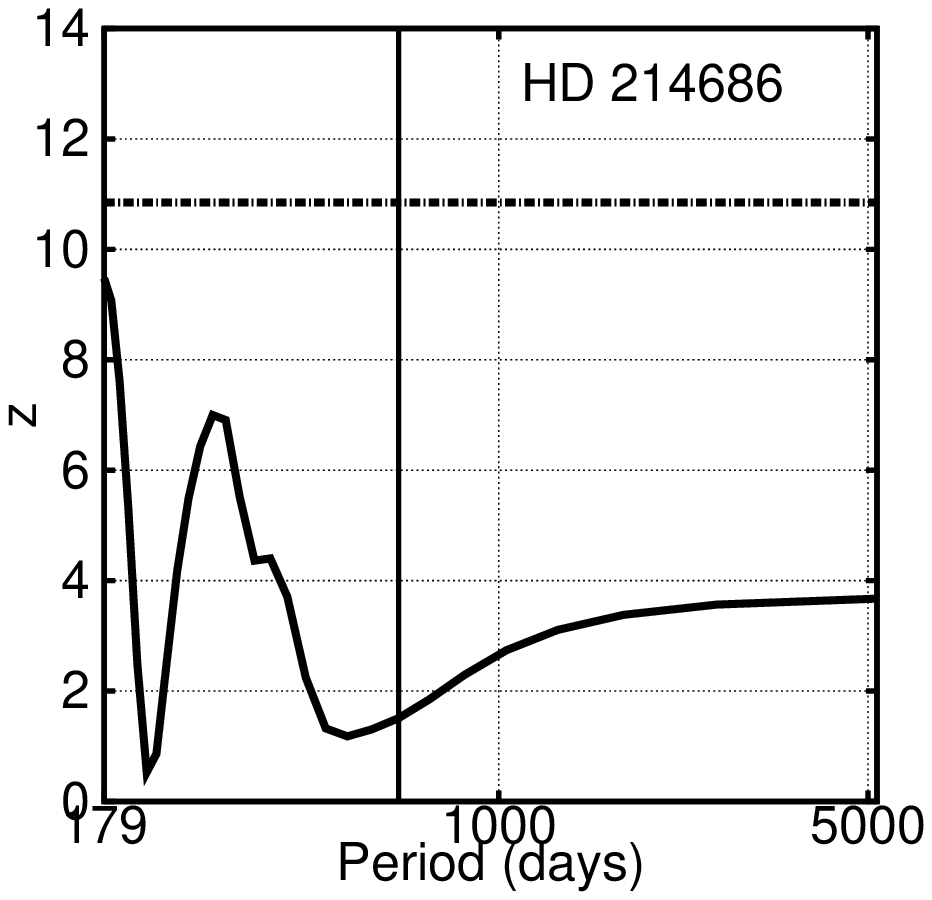}

\plottwo{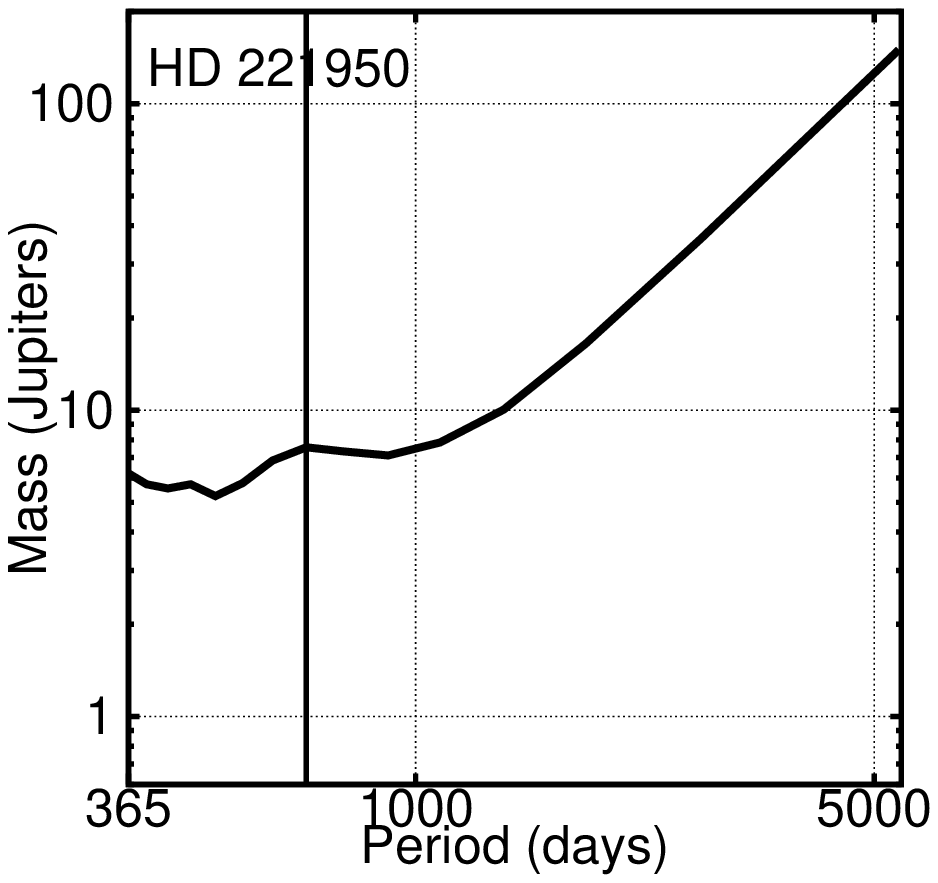}{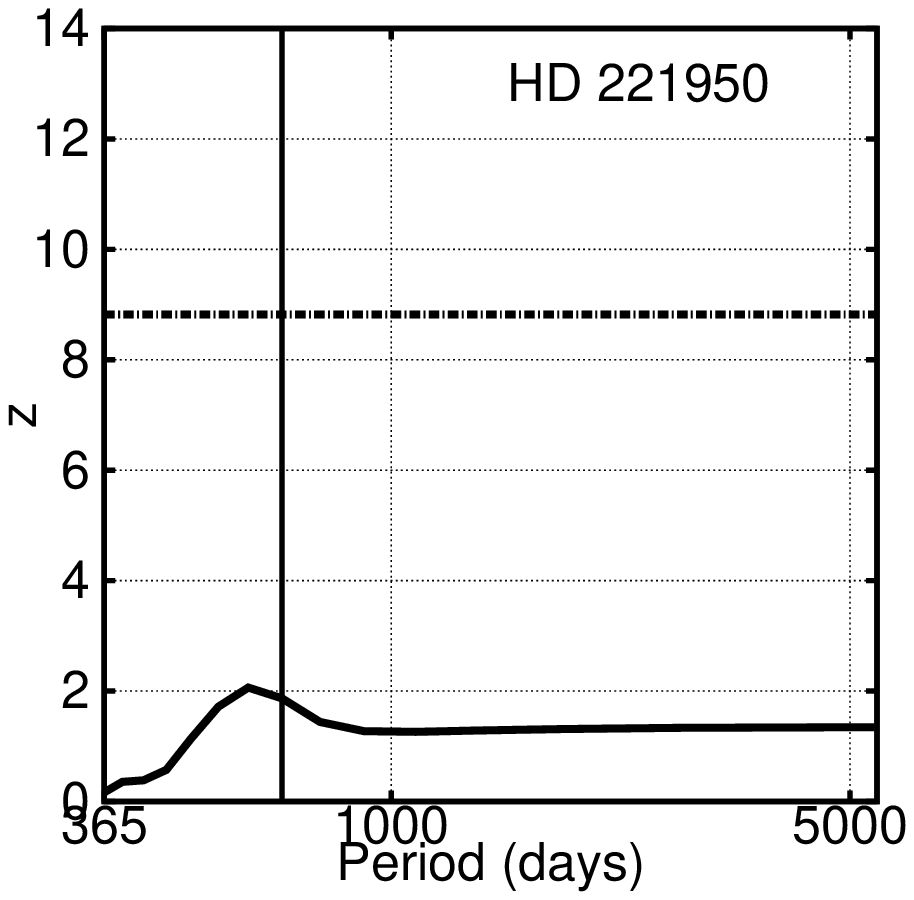}
\caption{Planet detection limits in the log P - log M space (orbital period
-mass; left panels) and periodograms (right panels). The solid line in the 
left panels is a planet detection limit corresponding to the 99\% confidence 
level. The vertical line near the orbital period of 1000 days denotes the
time span of the data set. 
\label{fig8}}
\end{figure}

\begin{figure}
\epsscale{0.85}
\plottwo{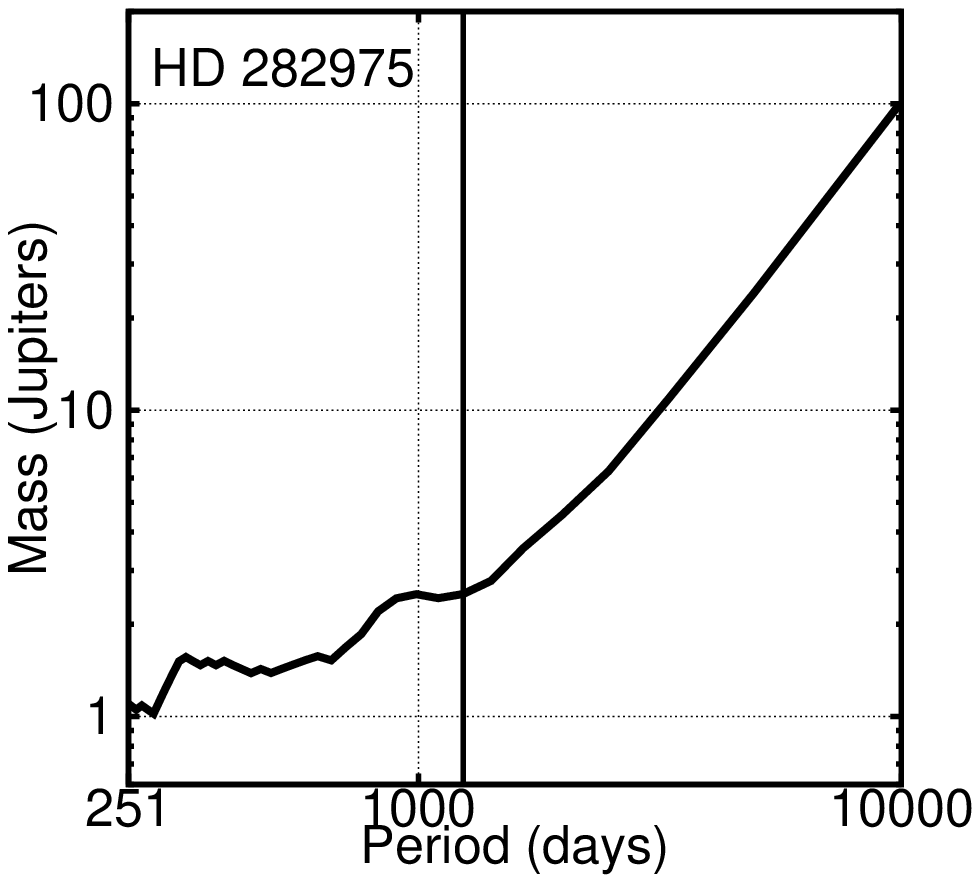}{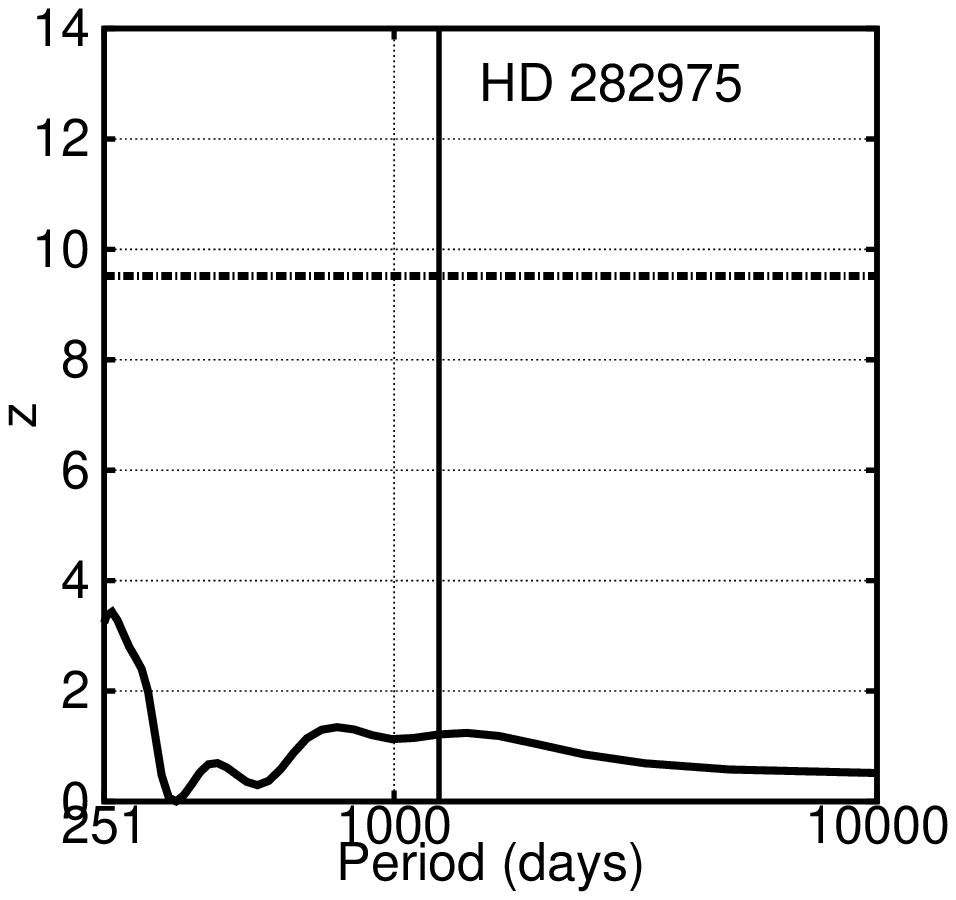}
\caption{Planet detection limits in the log P - log M space (orbital period
-mass; left panels) and periodograms (right panels). The solid line in the 
left panels is a planet detection limit corresponding to the 99\% confidence 
level. The vertical line near the orbital period of 1000 days denotes the
time span of the data set.
\label{fig9}}
\end{figure}

\begin{figure}
\epsscale{0.85}
\plottwo{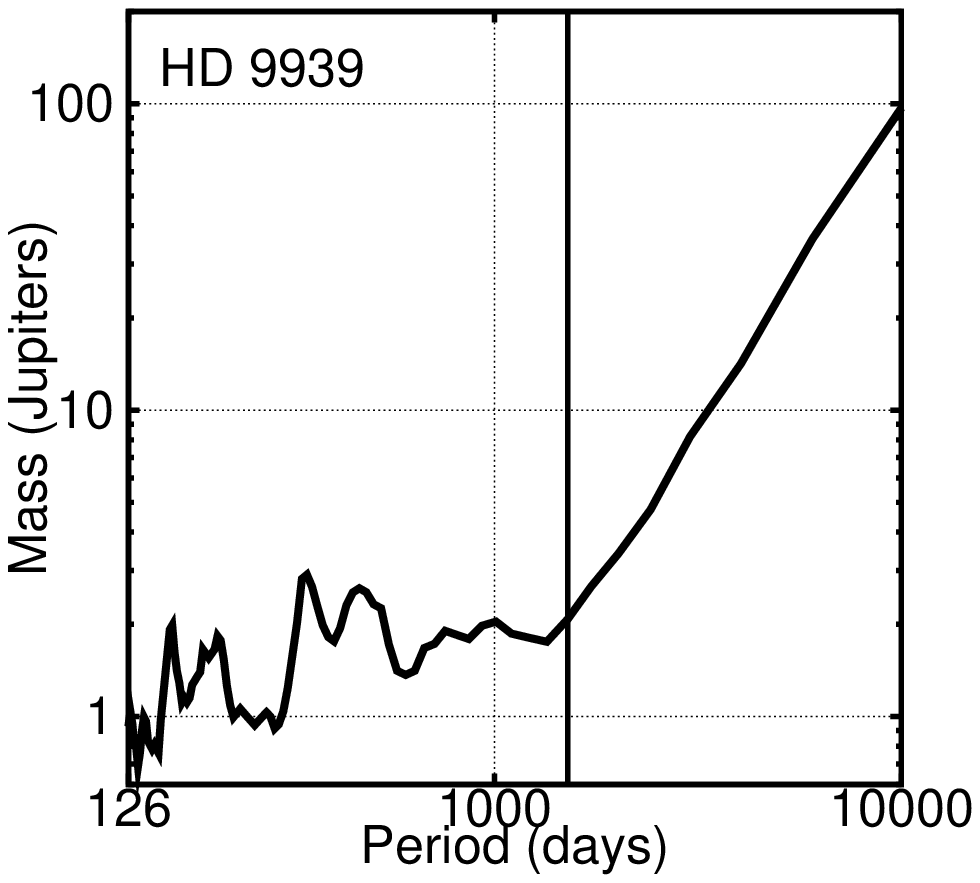}{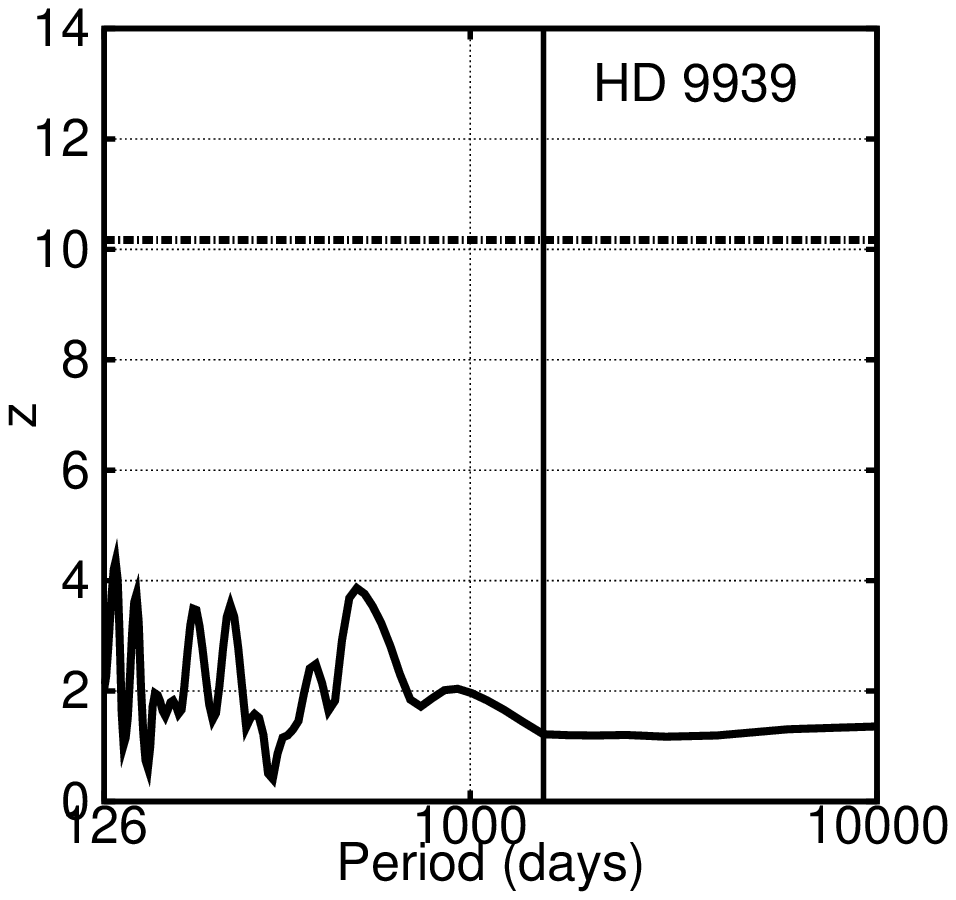}
\caption{Planet detection limits in the log P - log M space (orbital period
-mass; left panels) and periodograms (right panels) for HD9939
and planetary orbits with the eccentricities up to 0.6. The solid line in the left 
panels is a planet detection limit corresponding to the 99\% confidence level. 
The vertical line near the orbital period of 1000 days denotes the time span of 
the data set.
\label{fig10}}
\end{figure}

\clearpage

%
%

\begin{landscape}
\begin{deluxetable}{lrrrrrrrrrr}
\tabletypesize{\footnotesize}
\tablewidth{690pt}
\tablecaption{Targets and their radial velocity data sets}   
\tablehead{\colhead{\phm{Shane/CAT/Hamspec}} & \colhead{HD9939} &  
\colhead{HD13974} & \colhead{HD47415} & \colhead{HD78418} & 
\colhead{HD86146} & \colhead{HD195987} & 
\colhead{HD210027} & \colhead{HD214686} & \colhead{HD221950} & 
\colhead{HD282975}}
\startdata
V (mag)\dotfill & 6.99& 4.9& 6.38& 5.98& 5.12  & 
7.09& 3.76& 6.89& 5.70& 10.0  \\
Sp\dotfill &K0IV & G0V& F5V/F5V& G5IV-V& F5V/G0V  & 
G3V/K2V& F5V/G8V& F7V/F7V& F6V& G6V \\
r\dotfill & 6.3& 6.2& 1.7& 2.3& 2.5  &
6.7& 12& 1& 1.25&  1.1\\
M$_{1}$(M$_{\odot}$)\dotfill & 1.072\tablenotemark{a}& 0.6\tablenotemark{b}&
1.4\tablenotemark{c}& 1.15\tablenotemark{c}&  1.35\tablenotemark{d} & 
0.844\tablenotemark{e}&
1.326\tablenotemark{f}&   
1.25\tablenotemark{g}& 1.31\tablenotemark{g}& 1.0\tablenotemark{h} \\
M$_{2}$(M$_{\odot}$)\dotfill & 0.8383\tablenotemark{a}& 0.5\tablenotemark{b}& 
1.2\tablenotemark{c}& 1.0\tablenotemark{c}&  1.08\tablenotemark{d} &
0.665\tablenotemark{e}&
0.819\tablenotemark{f}&
1.25\tablenotemark{g}& 1.24\tablenotemark{g}& 0.9\tablenotemark{h} \\ 
P$_{orbital}$(d)\dotfill & 25.2& 10.0& 5.7& 19.4& 9.3  &
57& 10.2& 21.7& 45.5& 26 \\
a$_{stable}$(AU)\dotfill &0.61 & 0.23& 0.21& 0.61& 0.28  &
1.25& 0.29& 0.85& 1.37& 0.97 \\
P$_{stable}$(d)\dotfill & 126& 38& 22& 117& 35  &
416& 38& 179& 365& 251 \\
 & & & & & & & & & &  \\
All RVs & & & & &   & & & & &  \\
N$_{1}$+N$_{2}$\dotfill & 34& 28& 44& 50& 62  &
50& 146& 22& 22&  32\\
T$_{span}$(d)\dotfill & 1513& 663& 1695& 1695& 701  &
1474& 1925& 646& 681&  1238\\
rms$_{1}$(m$\,$s$^{-1}$)\dotfill & 19.5&22.5& 12.8& 11.5& 15.2  &
11.0& 17.2& 14.6& 48.4& 21.9 \\
rms$_{2}$(m$\,$s$^{-1}$)\dotfill & 36.0& 111.1& 25.6& 25.0&  67.6 &
48.0& 85.2& 14.7& 29.8& 12.5 \\
& & & & &   & & & & &  \\
Keck I/Hires & & & & &   & & & & &  \\ 
N$_{1}$+N$_{2}$\dotfill & 20& $\cdots$& 30& 26&  $\cdots$ &
22& 104& $\cdots$& $\cdots$&  32\\
rms$_{1}$(m$\,$s$^{-1}$)\dotfill & 6.8& $\cdots$& 12.8& 9.9&  $\cdots$ &
2.3& 16.5& $\cdots$& $\cdots$& 21.9 \\
rms$_{2}$(m$\,$s$^{-1}$)\dotfill & 21.4& $\cdots$& 10.6& 14.7&  $\cdots$ &
24.8& 89.0& $\cdots$& $\cdots$& 12.5 \\
$\sigma_{1}$(m$\,$s$^{-1}$)\dotfill & 1-4& $\cdots$& 4-11& 2-6&  $\cdots$ &
3-6& 5-32& $\cdots$& $\cdots$& 4-17 \\
$\sigma_{2}$(m$\,$s$^{-1}$)\dotfill & 4-10& $\cdots$& 5-11& 4-10&  $\cdots$ &
7-18& 12-65& $\cdots$& $\cdots$& 7-15 \\
$\epsilon_{1}$(m$\,$s$^{-1}$)\dotfill & 6& $\cdots$& 11& 5.5&  $\cdots$ &
0& 12& $\cdots$& $\cdots$&  20\\ 
$\epsilon_{2}$(m$\,$s$^{-1}$)\dotfill & 21& $\cdots$& 8& 4&  $\cdots$ &
20& 85& $\cdots$& $\cdots$&  11\\
& & & & &   & & & & &  \\
Shane/CAT/Hamspec & & & & &   & & & & &  \\
N$_{1}$+N$_{2}$\dotfill & 14& 28& 14& 24&  62 &
18& 30& 22& 22&  $\cdots$\\
rms$_{1}$(m$\,$s$^{-1}$)\dotfill & 14.0& 22.5& 12.3& 12.0&  15.2 &
14.2& 20.4& 14.6& 48.4&  $\cdots$\\
rms$_{2}$(m$\,$s$^{-1}$)\dotfill & 57.3& 111.1& 37.3& 31.0&  67.6 &
72.2& 85.5& 14.7& 29.8&  $\cdots$\\
$\sigma_{1}$(m$\,$s$^{-1}$)\dotfill & 4-14& 7-38& 5-21& 4-27&  6-34 &
7-17& 7-26& 11-26& 7-56& $\cdots$ \\
$\sigma_{2}$(m$\,$s$^{-1}$)\dotfill & 9-73& 13-91& 8-39& 6-35&  13-63 &
20-53& 16-100& 9-27& 8-55& $\cdots$ \\
$\epsilon_{1}$(m$\,$s$^{-1}$)\dotfill & 20& 19& 8& 11& 7 &
13& 18.5& 0& 48& $\cdots$ \\
$\epsilon_{2}$(m$\,$s$^{-1}$)\dotfill & 30& 110& 30& 37& 58 &
50& 68.0& 8& 32& $\cdots$ \\
 & & & & &   &  & & & &  \\
TNG/Sarg & & & & &   & & & & &  \\
N$_{1}$+N$_{2}$\dotfill & $\cdots$& $\cdots$& $\cdots$& $\cdots$&  $\cdots$ &
10& 12& $\cdots$& $\cdots$&  $\cdots$\\
rms$_{1}$(m$\,$s$^{-1}$)\dotfill & $\cdots$& $\cdots$& $\cdots$& $\cdots$& $\cdots$  &
5.1& 19.4& $\cdots$& $\cdots$& $\cdots$ \\
rms$_{2}$(m$\,$s$^{-1}$)\dotfill & $\cdots$& $\cdots$& $\cdots$& $\cdots$& $\cdots$  &
15.8& 89.9& $\cdots$& $\cdots$& $\cdots$ \\
$\sigma_{1}$(m$\,$s$^{-1}$)\dotfill & $\cdots$& $\cdots$& $\cdots$& $\cdots$& $\cdots$  &
6-12& 10-20& $\cdots$& $\cdots$&
$\cdots$ \\
$\sigma_{2}$(m$\,$s$^{-1}$)\dotfill & $\cdots$& $\cdots$& $\cdots$& $\cdots$& $\cdots$  &
11-17& 45-72& $\cdots$& $\cdots$&
$\cdots$ \\
$\epsilon_{1}$(m$\,$s$^{-1}$)\dotfill & $\cdots$& $\cdots$& $\cdots$& $\cdots$& $\cdots$  &
0& 12& $\cdots$& $\cdots$& $\cdots$
\\
$\epsilon_{2}$(m$\,$s$^{-1}$)\dotfill & $\cdots$& $\cdots$& $\cdots$& $\cdots$& $\cdots$ &
36& 38& $\cdots$& $\cdots$& $\cdots$
\enddata
\tablerefs{(a) \cite{Boden:06::}, (b) \cite{Hummel:95::}, (c)
\cite{Medeiros:99::}, (d) \cite{Batten:80::}, (e) \cite{Torres:02::}, 
(f) \cite{Boden:99a::}, (g) \cite{Tomkin:08::}, (h) \cite{Mermilliod:92::}}
\tablecomments{HD13974 is somewhat challenging to disentangle as 
the RV amplitudes are only 10.2 and 14.2 km$\,$s$^{-1}$ for the primary and 
secondary respectively. Its mass seems to be low by a factor of ~2, but we 
are using this published value anyways. HD221950 has relatively 
wide spectral lines compared to the remaining targets in this sample and this 
presumably has an impact on the tomographic disentangling and the final RV 
precision.}
\end{deluxetable}
\clearpage
\end{landscape}

\end{document}